\documentclass[
 reprint,
superscriptaddress,
 amsmath,amssymb,
 aps,
pra,
longbibliography
]{revtex4-1}

\usepackage{hyperref}
\usepackage{physics}
\usepackage{graphicx}
\usepackage{orcidlink}

\usepackage{array}

\newcommand{\e}{\mathrm{e}}
\newcommand{\diag}{\mathrm{diag}}
\renewcommand{\O}{\mathcal{O}}

\begin{document}

\title{Achieving quantum-limited sub-Rayleigh identification\\ of incoherent optical sources with arbitrary intensities}

\author{Danilo Triggiani\orcidlink{0000-0001-8851-6391}}
\email{Corresponding author: danilo.triggiani1@gmail.com}
\affiliation{Dipartimento Interateneo di Fisica, Politecnico di Bari, 70126, Bari, Italy}
\affiliation{Istituto Nazionale di Fisica Nucleare (INFN), Sezione di Bari, 70126 Bari, Italy}

\author{Cosmo Lupo\orcidlink{0000-0002-5227-4009}}
\affiliation{Dipartimento Interateneo di Fisica, Politecnico di Bari, 70126, Bari, Italy}
\affiliation{Dipartimento Interateneo di Fisica, Universit\`a di Bari, 70126, Bari, Italy}
\affiliation{Istituto Nazionale di Fisica Nucleare (INFN), Sezione di Bari, 70126 Bari, Italy}

\date{\today}

\begin{abstract}

The Rayleigh diffraction limit imposes a fundamental restriction on the resolution of direct imaging systems, hindering the identification of incoherent optical sources, such as celestial bodies in astronomy and fluorophores in bioimaging. Recent advances in quantum sensing have shown that this limit can be circumvented through spatial demultiplexing (SPADE) and photon detection, i.e. a semi-classical detection strategy.
However, the general optimality for arbitrary intensity distributions and bright sources remains unproven. In this work, we develop a general model for incoherent light with arbitrary intensity undergoing diffraction. We employ this framework to compute the quantum Chernoff exponent for generic incoherent-source discrimination problems, focusing on the sub-diffraction regime. We show that, surprisingly, SPADE measurements saturate the quantum Chernoff bound only when certain compatibility conditions are met. 
These findings suggest that collective measurements may actually be needed to achieve the ultimate quantum Chernoff bound for the discrimination of specific incoherent sources.
For the fully general case, our analysis can still be used to find the best SPADE configurations, generally achieved through a rotation of the SPADE interferometer that depends on the discrimination task. 
We also simulated the efficiency of a simplified Bayesian test that we developed for this identification task and show that the saturation of the Chernoff bound is already achieved for a finite number of repetitions $N\leqslant 5000$.
Our results advance the theory of quantum-limited optical discrimination,  with possible applications in diagnostics, automated image interpretation, and galaxy identification.
\end{abstract}

\maketitle

\section{Introduction}

The blurring effect caused by the optical diffraction of light passing through small apertures is the principal cause of loss of detail and resolution when imaging incoherent sources.
Dubbed the Rayleigh criterion or diffraction limit, for more than a century this limitation was thought to be an unavoidable downside of imaging techniques, requiring the use of larger apertures to increase angular resolution for a fixed wavelength~\cite{Rayleigh1879}.
However, modern theoretical tools developed within the quantum sensing formalism, such as the quantum Fisher information~\cite{Helstrom1969}, painted a different picture:
the diffraction limit is not ubiquitously present in every imaging technique, but instead is a specific drawback of direct imaging~\cite{Tsang2016L, Lupo2016, Rehacek2017A, Yu2018, Tsang2019A, Zhou2019, Napoli2019}. 
As such, it can be overcome with more clever choices of measurement schemes.
Indeed, several alternative approaches that achieve a higher resolution of thermal sources than direct imaging have been developed, such as spatial demultiplexing (SPADE), which has been proposed for the estimation of the distances of point sources in one dimension~\cite{Tsang2016X, Rehacek2017} and two dimensions~\cite{Ang2017}, of the second or higher moments of intensity distributions~\cite{Tsang2017, Tsang2018}, of the transversal lengths of a sources~\cite{Dutton2019}, of the distance and centroid of sources simultaneously~\cite{Hervas2024}, and of the the location of multiple sources~\cite{Bisketzi2019}, while experimental demonstrations~\cite{Pushkina2021, Frank2023, Santamaria2023, Wallis2025}, and analyses to tame experimental imperfections~\cite{Sorelli2021, Linowski2023} have also been recently performed. 
Alternatives to SPADE include homodyne and heterodyne detection~\cite{Yang2016, Yang2017}, super-resolved localization via image inversion (SLIVER)~\cite{Tsang2016L}, super-resolved position localization via inversion of coherence (SPLICE)~\cite{Tham2017, Bonsma-Fisher2019}, pump-driven non-linear optical processes~\cite{Darji2024} or two-photon interference~\cite{Parniak2018, Muratore2025} (see also Refs.~\cite{Tsang2019, Defienne2024} for comprehensive reviews).
The overarching operating principle of these techniques is to manipulate the optical field in such a way that the signal and background noise are separated and sorted into different modes~\cite{Tsang2019}.

Incoherent source discrimination is an example of tasks that is severely hindered by the diffraction limit when performed through direct imaging.
For this type of problem, the goal is to find the optimal measurement and hypothesis test that identify a light-emitting object within a pool of possible hypotheses~\cite{Helstrom1969}.
Initiated by studies of quantum detection theory applied to point sources~\cite{Helstrom1973}, and bolstered by the development of benchmarks for the error probability, such as the quantum Chernoff bound~\cite{Audenaert2007}, this line of research has produced a plethora of interesting results and proposals in the last decade.
SPADE was shown to be successful (and indeed optimal) in one-versus-two point-sources discrimination, both theoretically~\cite{Lu2018, Huang2021, Jha2025} and experimentally~\cite{Zanforlin2022, Santamaria2024}.
At the same time, great effort has been put into the realization of hypothesis tests that are less susceptible to experimental imperfections, such as crosstalk and dark counts~\cite{Gessner2020, DeAlmeida2021, Schlichtholz2024, Linowski2025}, that take into account partial coherence~\cite{Zhang2025}, or that employ alternative sub-Rayleigh schemes, such as SLIVER~\cite{Zhang2024}.
Recently, Grace and Guha provided an interesting analysis of quantum-optimal measurements for faint-source discrimination with more generic intensity distributions~\cite{Grace2022}.
It was found that the SPADE approach is still optimal for this task, albeit commutativity between the intensity distribution covariance matrices was assumed, rendering the results not fully generic.
SPADE was also recently shown to be effective when combined with machine learning techniques for more generic faint source quantum classification tasks~\cite{Buonaiuto2025}.

However, comprehensive results related to subdiffraction discrimination under more generic conditions are still scarce. 
First and foremost, the single-photon assumption becomes restrictive for intense sources, i.e.~in the microwave to far-infrared regimes or for bright stars and fluorophores. 
Moreover, it is still unclear whether SPADE techniques are universally optimal in the subdiffraction regime for generic intensity distribution discrimination and whether the optimal SPADE is hypothesis dependent (see Table 1 for a ‌summary of the literature).
In this work, we address both these open questions.
First, we develop a Gaussian model for incoherent light with arbitrary intensity collected by passive and linear optics that induces a point-spread function (PSF).
Due to its Gaussian nature, our model can be easily employed in theoretical analyses of incoherent light affected by diffraction beyond the purpose of detection theory~\cite{Weedbrook2012}.
We then employ our model to analyze the quantum discrimination problem of arbitrary incoherent sources.
We evaluate the quantum Chernoff exponent in the general scenario, and we find its expression for the specific sub-cases of faint sources and commuting covariance matrices.
We then specialize our model to the subdiffraction regime for Gaussian PSFs, where the main and most surprising results of our analysis can be found.

\begin{table}[t]
    \centering
    \includegraphics[width=.9\columnwidth]{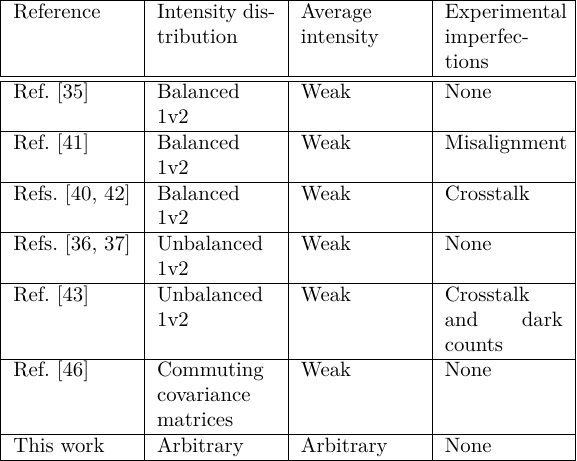}
    \caption{Summary table of some recent theoretical results for incoherent source discrimination using spatial demultiplexing.}
    \label{tab:placeholder}
\end{table}

By comparing the Chernoff exponent associated with SPADE measurements with the ultimate benchmark given by the quantum Chernoff exponent, we find that SPADE is \textit{not} always optimal, in the sense that it does not always saturate the quantum Chernoff bound.
Instead, optimality is achieved only if some compatibility conditions are met by the intensity distribution associated with the two hypotheses, such as, but not limited to, the commutativity between the covariance matrices of the two sources.
If these conditions are not met, achieving the ultimate quantum Chernoff bound may require collective detection strategies that fully exploit the quantum nature of the optical field, such as entanglement, photon counting or projection on multi-photon states~\cite{Conlon2023}.
Nevertheless, our analysis shows how to numerically find the best SPADE measurement in the subdiffraction regime, which, in the most generic scenario, we find to be the projection over the Hermite-Gauss mode $00$ and a specific hypotheses-dependent rotation of the Hermite-Gauss modes $01$ and $10$. 
This measurement can be achieved by simply physically rotating the demultiplexing interferometer.
We finally develop and numerically simulate a Bayesian hypotheses test based on counting the number of frames in which a photon has been detected in the rotated Hermite-Gauss mode $01$, in $10$, or neither of them.
Interestingly this test does not require in principle photon-resolving detectors, as on-off detectors suffice for the task.
We find in our simulations that this test achieves the classical Chernoff bound exponential decay of the error probability and a $2\sigma=95\%$ confidence of success already for $N\leqslant 5000$, with this number drastically decreasing for more intense sources.

Our result showcases an important advancement in the field of incoherent-sources subdiffration discrimination, outlining the optimal strategies also for generic intensity distribution, with possible applications, among others, in automated diagnostics~\cite{Farahani2016} and galaxy identification~\cite{Robertson2023}.

\section{Exact model for arbitrary intensities}

The light emitted by an incoherent source with a given intensity distribution $I_{i,j}=I(x_i,y_j)\delta x\delta y$, here expressed in discrete notation for simplicity, with $\delta x$ and $\delta y$ the dimensions of the pixels, can be generally expressed as tensor products of thermal states $\rho_0 = \bigotimes_{i,j}\rho_{i,j}$, with
\begin{equation}
\begin{gathered}
\rho_{i,j}=\int\dd^2\alpha_{i,j}\ P_{i,j}(\alpha_{i,j})\ketbra{\alpha_{i,j}},\\
P_{i,j}(\alpha_{i,j})=\frac{1}{\pi I_{i,j}} \, \e^{-\frac{\abs{\alpha_{i,j}}^2}{I_{i,j}}},
\end{gathered}
\label{eq:ThermalMulti}
\end{equation}
where $\{\ket{\alpha_{i,j}}\}$ is the coherent state basis for the $(i,j)$th pixel, and $P_{i,j}
(\cdot)$ is the Glauber-Sudarshan P-function of the state $\rho_{i,j}$~\cite{MandelWolf1995}.
Comparing with the generic P-function of a $M$-mode thermal state $P(\alpha)=\det(\pi\Gamma)^{-1/2}\exp(-\alpha^T\Gamma^{-1}\alpha)$, with $\alpha\equiv(\alpha_{r1},\alpha_{i1},\dots,\alpha_{rM},\alpha_{iM})$, $\alpha_{rk}^2+\alpha_{ik}^2=\abs{\alpha_k}^2$ and $\Gamma$ the $2M\times2M$ covariance matrix, we can see that Eq.~\eqref{eq:ThermalMulti} corresponds to the choice of uncorrelated sources, with covariance matrix at the source plane $\Gamma_S=\gamma_S\otimes I_2$ with $\gamma_S=\diag(\{I_{i,j}\})$.
Propagating through a linear and passive optical system, the state $\rho_0$ transforms according to the unitary map 
\begin{equation}
\mathcal{U}\left(a^\dag_{i,j}\right)
= \sqrt{1-\nu} \,  e_{i,j}^\dag+\sqrt\nu \,  \sum_{i'.j'}C_{\substack{i,j\\i',j'}}c^\dag_{i',j'},
\end{equation}
where $c_{i,j}$ denotes the observed output optical mode in pixel $i,j$, $\nu$ is the transmission coefficient of the optical system, $C_{\substack{i,j\\i',j'}}=\psi(x_{i'}-x_i)\psi(y_{j'}-y_j)$ is the scattering matrix that is determined by the PSF 
$\Psi(x,y)=\psi(x)\psi(y)$, while $e_{i,j}$ are $M$ auxiliary environment modes introduced to preserve the unitarity.
Being the coherent states $\ket{\alpha_{i,j}}$ eigenstates of the annihilation operator $a_{i,j}$, the same transformation $\mathcal{U}$ can be thought to be applied to the coherent field amplitude $\alpha_{i,j}$ 
(see Appendix~\ref{app:Model} for a formal approach).
Moreover, linear operations preserve the Gaussian nature of the P-function of $\rho_0$.
As a further simplification, we will assume that the PSF $\psi$ is real and symmetric, so that it does not mix real and imaginary parts of the coherent field amplitudes associated with the observed modes $c_{i,j}$, thus preserving the Kronecker product structure of $\Gamma=\gamma\otimes I_2$, with $\gamma=(C\gamma_S C)$.
After tracing out the environment modes, which corresponds to consider the covariance sub-matrix associated to the modes $c_{i,j}$, the thermal state $\rho$ in the image plane can thus be written as
\begin{equation}
\begin{gathered}
\rho=\int \left(\prod_{l,m}\dd^2\alpha_{l,m}\right) P(\{\alpha_{l,m}\})\left(\bigotimes_{l,m}\ketbra{\alpha_{l,m}}{\alpha_{l,m}}\right),\\
P(\{\alpha_{l,m}\})=\frac{1}{\det(\pi\gamma)}\e^{-\boldsymbol{\alpha}^\dag\gamma^{-1}\boldsymbol{\alpha}}
\label{eq:ThermalState1}
\end{gathered}
\end{equation}
with 
\begin{equation}
\begin{gathered}
\gamma_{\substack{l,m\\l',m'}}=\nu\sum_{i,j} I_{i,j}\psi(x_i-x_l)\psi(x_i-x_{l'})\psi(y_j-y_m)\psi(y_j-y_{m'}),
\end{gathered}
\label{eq:CovMatGen}
\end{equation}
where $I_0=\nu\sum_{i,j}I_{i,j}$ is the total intensity of the source.
In the remainder of this work, we will absorb the transmission coefficient $\nu$ in the intensity distribution $\nu I_{i,j}\rightarrow I_{i,j}$.

\section{Optimal object identification}

Our goal is to discriminate between two incoherent sources $\rho_i$ with intensity distributions $I_{i}(x,y)$, $i=1,2$, with same total average intensity $I_0$, by detecting the emitted light and formulating a test that guesses which of the two hypotheses is true.
After collecting light for a number $N\gg1$ of repetitions, any hypothesis test has an error probability $P_e$ that is bounded from below by the quantum Chernoff bound $P_e\geqslant \exp(-N\xi_{Q})$, where $\xi_{Q}$ is the quantum Chernoff exponent~\cite{Audenaert2007}
\begin{equation}
	\xi_Q=-\ln\min_{s\in[0,1]}\Tr(\rho_1^s\rho_2^{1-s}),
\end{equation}
where Tr denotes the trace on operators.
For Gaussian states $\rho_1,\rho_2$ with first momenta equal to zero, such as thermal states shown in Eq.~\eqref{eq:ThermalState1}, we show in Appendix~\ref{app:QCB} that the quantum Chernoff exponent can be written as
\begin{equation}
    \xi_Q=-\ln\min_{s\in [0,1]}\left(2^M \frac{\det(g_s(\gamma_1))\det(g_{1-s}(\gamma_2))}{\det(\lambda_s(\gamma_1)+\lambda_{1-s}(\gamma_2))}\right),
    \label{eq:GaussianQCB}
\end{equation}
with $M$ the number of pixels, $g_s(x)=[(x+1)^s-x^s]^{-1}$, $\lambda_s(x)=[(x+1)^s+x^s]/[(x+1)^s-x^s]$, where $\lambda_s(\gamma_i),g_s(\gamma_i)$ are defined through the eigendecompositions of $\gamma_i=U_iD_iU_i^T$, i.e.~$f_s(\gamma_i)=U_if_s(D_i)U_i^T$, for $f=g,\lambda$ and $i=1,2$.
Equation~\eqref{eq:GaussianQCB} is general and holds independently of the intensity of the source and the strength of the diffraction introduced by the optical system.
The high dimensionality of the covariance matrices $\gamma_i$ renders the analytical evaluation of $\xi_Q$ in Eq.~\eqref{eq:GaussianQCB} generally intractable.
However, it is possible to simplify its expression under some specific conditions.
In the remainder of this section, we will retrieve simple expressions for the quantum Chernoff bound for the weak-source regime and for commuting covariance matrices.
Section~\ref{sec:subd} will be fully dedicated to the subdiffraction regime.

\subsection{Faint sources regime}

From Eq.~\eqref{eq:CovMatGen}, we see that the covariance matrix $\gamma_i$ is proportional to the overall intensity $I_0$ of the source, assuming to keep $I_{i,j}/I_0$ fixed, hence also its eigenvalues are proportional to $I_0$.
We can use this property to expand the functions $g_s(x)$ and $\lambda_s(x)$ in orders of $x^s$ up to the linear term in $x$.
After some careful steps in the calculation of $\xi_Q$ expanded up to the linear order in $I_0$ (see Appendix~\ref{app:Faint}), we get
\begin{equation}
	\xi_Q\overset{I_0\ll1}{=}I_0-\min_{s\in[0,1]}\tr{\gamma_1^s\gamma_2^{1-s}},
    \label{eq:QCBWeak}
\end{equation}
where tr denotes the matrix trace operation.
This result agrees with previous literature where, in the faint source regime, the quantum Chernoff coefficient $\xi_Q\simeq I_0(1-\e^{-\xi_Q^{(1)}})=I_0(1- \mathrm{Tr}[(\rho_1^{(1)})^{s}(\varrho_2^{(1)})^{1-s}])$ is expressed in terms of the "per-photon" quantum Chernoff coefficient $\xi_Q^{(1)}$ and the "per-photon" quantum states $\rho_{1/2}^{(1)}$~\cite{Grace2022}.
Indeed, in the regime of small $I_0$, a thermal state of the type of Eq.~\eqref{eq:ThermalState1} can be expanded up to order $I_0$ in the Fock basis, retrieving 
\begin{equation}
	\rho=(1-\tr{\gamma})\ketbra{0}{0} + \sum_{l,m}\gamma_{\substack{l,m\\l',m'}}\ketbra{1_{l,m}}{1_{l',m'}}+\O(I_0^2)
\end{equation}
where $\rho^{(1)}=\frac{1}{I_0}\sum_{l,m}\gamma_{\substack{l,m\\l',m'}}\ketbra{1_{l,m}}{1_{l',m'}}$ coincides with the normalized covariance matrix $\gamma/I_0$.
Notice that Eq.~\eqref{eq:QCBWeak} is valid also for non-commuting covariance matrices.

\subsection{Commuting covariance matrices}

If $\gamma_1$ and $\gamma_2$ commute, they share a common eigenvector basis $U_1=U_2=U$ so that $\gamma_i=UD_iU^T$, and Eq.~\eqref{eq:GaussianQCB} can be written exclusively via the eigenvalues $D_i$ of the covariance matrices. 
We can thus evaluate, for each corresponding eigenvalue $x_i$ and $y_i$ of $\gamma_1$ and $\gamma_2$, the expression
\begin{equation}
\frac{g_s(x_i)g_{1-s}(y_i)}{\lambda_s(x_i)+\lambda_{1-s}(y_i)}=\frac{1}{2}\frac{1}{(1+x_i)^s(1+y_i)^{1-s}-x_i^sy_i^{1-s}}	
\end{equation}
so that we have
\begin{equation}
	\xi_Q = \ln\max_{s\in[0,1]}\det((1+\gamma_1)^s(1+\gamma_2)^{1-s}-\gamma_1^s\gamma_2^{1-s}).
    \label{eq:QCBCommute}
\end{equation}
Moreover, since the commutativity of the covariance matrices implies the commutativity of the density matrices, the quantum Chernoff coeffcient in Eq.~\eqref{eq:QCBCommute} can always be achieved by the measurement that projects on the common eigenbasis of the density matrices, that is, the Fock basis rotated by $U$.

\section{Subdiffraction regime}\label{sec:subd}

Here we discuss the most interesting regime of subdiffraction source discrimination.
In this regime, the width of the PSF $\psi$ is larger than or comparable to the size of the imaged source of light, so that direct imaging is renownedly incapable to successfully discriminate between different sources.
In the following, we first adapt our generic description of incoherent sources to the subdiffraction regime, then evaluate the quantum Chernoff exponent $\xi_Q$, and we compare it with the Chernoff exponent associated with SPADE measurement in different physical subcases of interest.

\subsection{Subdiffraction model for arbitrary intensities}
We want to specify the expressions for the thermal state and its covariance matrix in Eqs.~\eqref{eq:ThermalState1} and~\eqref{eq:CovMatGen} to the subdiffraction scenario.
For simplicity, in this section we will consider Gaussian PSFs with variance $\sigma^2$. 
The generalization to any PSF is rather straightforward and ultimately simply introduces different coefficients due to the different PSF autocorrelation functions. We refer to the work of Grace and Guha for the case of generic PSF in the single-photon regime~\cite{Grace2022}.

First, we notice that it is possible to change the expression of the covariance matrix $\gamma$ in Eq.~\eqref{eq:CovMatGen} from the position basis to a different basis $\alpha_{s,t}$ through the relation
\begin{equation}
\gamma^\alpha_{\substack{s,t\\s't'}}=\sum_{\substack{l,m\\l',m'}}\braket{\alpha_{s,t}}{x_l,y_m}\gamma_{\substack{l,m\\l',m'}}\braket{x_{l'},y_{m'}}{\alpha_{s',t'}}.
\end{equation}
We can then choose the new basis $\ket{\alpha_{s,t}}=\ket{HG_s^x,HG_t^y}$ to be the 2D Hermite-Gauss basis, as we will see it is the natural choice to express $\gamma$ in the subdiffraction regime.
Recovering the continuous notation $\gamma_{\substack{l,m\\l',m'}}\rightarrow\gamma(x_1,y_1;x_2,y_2)$, we can write
\begin{multline}
\gamma^{HG}_{\substack{s,t\\s't'}}=\int\dd x\dd y\dd x_1\dd x_2\dd y_1\dd y_2\ 
\gamma(x_1,y_1;x_2,y_2)\times\\
\times\phi_s(x_1)\phi_{s'}(x_2)\phi_t(y_1)\phi_{t'}(y_2) \\
=\int\dd x\dd y\ I(x,y) \frac{\e^{-\frac{x^2+y^2}{4\sigma^2}}}{\sqrt{s!s'!t!t'!}}\left(\frac{x}{2\sigma}\right)^{s+s'}\left(\frac{y}{2\sigma}\right)^{t+t'},
\label{eq:CovHG}
\end{multline}
where $\phi_s(x)=\braket{x}{HG_s^{x}} = \frac{\mathcal{N}^{1/2}}{\sqrt{\sqrt{2\pi\sigma^2}2^s s!}} \, e^{-\frac{x^2}{4\sigma^2}} H_s\left( \frac{x}{\sqrt{2}\sigma} \right)$, and $H_s$ denotes the $s$th physicist's Hermite polynomial.
If we rewrite the integral in Eq.~\eqref{eq:CovHG} in terms of the dimensionless quantities $\bar{I}(\bar{x},\bar{y})=\theta^2 I(\theta \bar{x},\theta\bar{y})/I_0$, with $\theta^2$ ``size" of the image and $I_0$ its total intensity, we can expand $\gamma^{HG}$ in orders of $\chi=\theta/(2\sigma)$ up to terms of order $\chi^2$. 
We obtain
\begin{equation}
\resizebox{.95\columnwidth}{!}{$
\gamma^{HG}=I_0\begin{pmatrix}
1-\chi^2(m^{2,0}+m^{0,2}) & \chi m^{1,0} & \chi m^{0,1} & \frac{\chi^2}{\sqrt{2}}m^{2,0} & \chi^2m^{1,1} & \frac{\chi^2}{\sqrt{2}}m^{0,2}\\
\chi m^{1,0} & \chi^2m^{2,0} & \chi^2 m^{1,1} & 0 & 0 & 0\\
\chi m^{0,1} & \chi^2 m^{1,1} & \chi^2 m^{0,2} & 0 & 0 & 0\\
\frac{\chi^2}{\sqrt{2}}m^{2,0} & 0 & 0 & 0 & 0 & 0\\
\chi^2 m^{1,1} & 0 & 0 & 0 & 0 & 0\\
\frac{\chi^2}{\sqrt{2}}m^{0,2} & 0 & 0 & 0 & 0 & 0
\end{pmatrix}
$}
\end{equation}
with
\begin{equation}
m^{s,t}=\int\dd x\dd y\ \bar{I}(x,y) x^s y^t
\end{equation}
$(s,t)$-th moment of the dimensionless intensity distribution $\bar{I}(x,y)$.
Notice that the only observable correlations at the order $\chi^2$ are the ones between the modes $\langle HG_{st}HG_{s't'}\rangle$ with $s+s'+t+t'\leqslant2$, corresponding to non-vanishing entries of $\gamma^{HG}$.

\subsection{Quantum Chernoff bound}

To define an interesting subdiffraction discrimination problem, we assume that the two hypotheses involve intensity distributions that are centered at the same point, that we set as the origin of the reference frame, so that $m_i^{1,0}=m_i^{0,1}=0$ for $i=1,2$.
In this way the source cannot be easily recognized simply finding its center through direct imaging.
In this scenario, the covariance matrices share the same block diagonal structure $\gamma_i=\gamma_i^\alpha\oplus\gamma_i^\beta$, $i=1,2$, with
\begin{gather}
\resizebox{.95\columnwidth}{!}{$
	\gamma_i^\alpha=I_0
    \begin{pmatrix}
	    1-\chi^2(m_i^{2,0}+m_i^{0,2}) & \frac{\chi^2}{\sqrt{2}}m_i^{2,0} & \chi^2 m_i^{1,1} & \frac{\chi^2}{\sqrt{2}}m_i^{0,2}\\
        \frac{\chi^2}{\sqrt{2}}m_i^{2,0} & 0 & 0 & 0\\
        \chi^2 m_i^{1,1} & 0 & 0 & 0\\
        \frac{\chi^2}{\sqrt{2}}m_i^{0,2} & 0 & 0 & 0
	\end{pmatrix}
    $},\\
    \gamma_i^\beta=I_0\chi^2
    \begin{pmatrix}
        m_i^{2,0} & m_i^{1,1}\\
        m_i^{1,1} & m_i^{0,2}
    \end{pmatrix},
\end{gather}
so that the density matrices $\rho_i=\rho_i^\alpha\otimes\rho_i^\beta$ can be factorized.
The same block-diagonal structure is preserved for $g_s(\gamma_i)=g_s(\gamma_i^\alpha)\oplus g_s(\gamma_i^\beta)$ and $\lambda_s(\gamma_i)=\lambda_s(\gamma_i^\alpha)\oplus\lambda_s(\gamma_i^\beta)$ in Eq.~\eqref{eq:GaussianQCB}, so that we can simplify the calculation of the quantum Chernoff exponent
\begin{equation}
	\xi_Q=-\ln\min_{s\in [0,1]}\left(2^6 \prod_{\zeta=\alpha,\beta}\frac{\det(g_s(\gamma^\zeta_1))\det(g_{1-s}(\gamma^\zeta_2))}{\det(\lambda_s(\gamma^\zeta_1)+\lambda_{1-s}(\gamma^\zeta_2))}\right).
\end{equation}
We notice that $\gamma_i^\alpha$ has only one nonzero eigenvalue up to order $\chi^2$. 
We show in Appendix~\ref{app:Subdiff} that
\begin{equation}
	2^4\frac{\det(g_s(\gamma^\alpha_1))\det(g_{1-s}(\gamma^\alpha_2))}{\det(\lambda_s(\gamma^\alpha_1)+\lambda_{1-s}(\gamma^\alpha_2))}=1 + \O(\chi^4),
    \label{eq:IndistinguibleAlpha}
\end{equation}
namely, the quantum Chernoff bound is independent of $\rho_i^\alpha$ up to the order $\chi^2$.
Moreover, since $\gamma_i^\beta\propto\chi^2$, similar steps to the ones performed for $I_0\ll 1$ in Appendix~\ref{app:Faint} can be done to retrieve
\begin{equation}
	\xi_Q \overset{\chi\ll1}{=}\max_{s\in [0,1]}\left[s\tr[\gamma^\beta_1]+(1-s)\tr[\gamma^\beta_2]-\tr[\gamma^{\beta\,s}_1\gamma^{\beta\,1-s}_2]\right].
\end{equation}
It is now convenient to write the matrices $\gamma^\beta_{i}$ through their spectral decomposition $\gamma_i^\beta=U_i^\beta D_i^\beta U_i^{\beta T }$, where $D_i^\beta=I_0\chi^2\diag(V_{ix},V_{iy})$, while $U_i^\beta\equiv U(\theta_i)$, with
\begin{equation}
	U(\theta)=\begin{pmatrix}
	    \cos\theta & -\sin\theta\\
        \sin\theta & \cos\theta
	\end{pmatrix} \, ,
\end{equation}
represents a image rotation of an angle $\theta$.
In this notation we can write $\mathrm{tr}[\gamma^{\beta\,s}_1\gamma^{\beta\,1-s}_2]=\sum_{i,j=1}^2 U(\Delta\theta)^2_{i,j} (D_1^\beta)^s_i(D_2^\beta)^{1-s}_j$, with $\Delta\theta=\theta_1-\theta_2$, and finally
\begin{multline}
	\xi_Q \overset{\chi\ll1}{=}I_0\chi^2\max_{s\in [0,1]}\Big[s(V_{1x}+V_{1y})+(1-s)(V_{2x}+V_{2y})\\
    -\cos^2\Delta\theta( V_{1x}^sV_{2x}^{1-s}+V_{1y}^sV_{2y}^{1-s})-\sin^2\Delta\theta( V_{1x}^sV_{2y}^{1-s}+V_{1y}^sV_{2x}^{1-s})\Big].
    \label{eq:QCB}
\end{multline}
This is the generic expression of the Chernoff exponent associated with a discrimination task between two incoherent sources with arbitrary intensities in the subdiffraction regime.
We can see that $\xi_Q$ depends on the principal variances $V_{ix/y}$ of the intensity distributions of the two sources and on the angle $\Delta\theta$ between the directions of the principal variances.

\subsection{Chernoff bound for SPADE}
\label{sec:CB}

\begin{figure}
    \centering
    \includegraphics[width=0.95\linewidth]{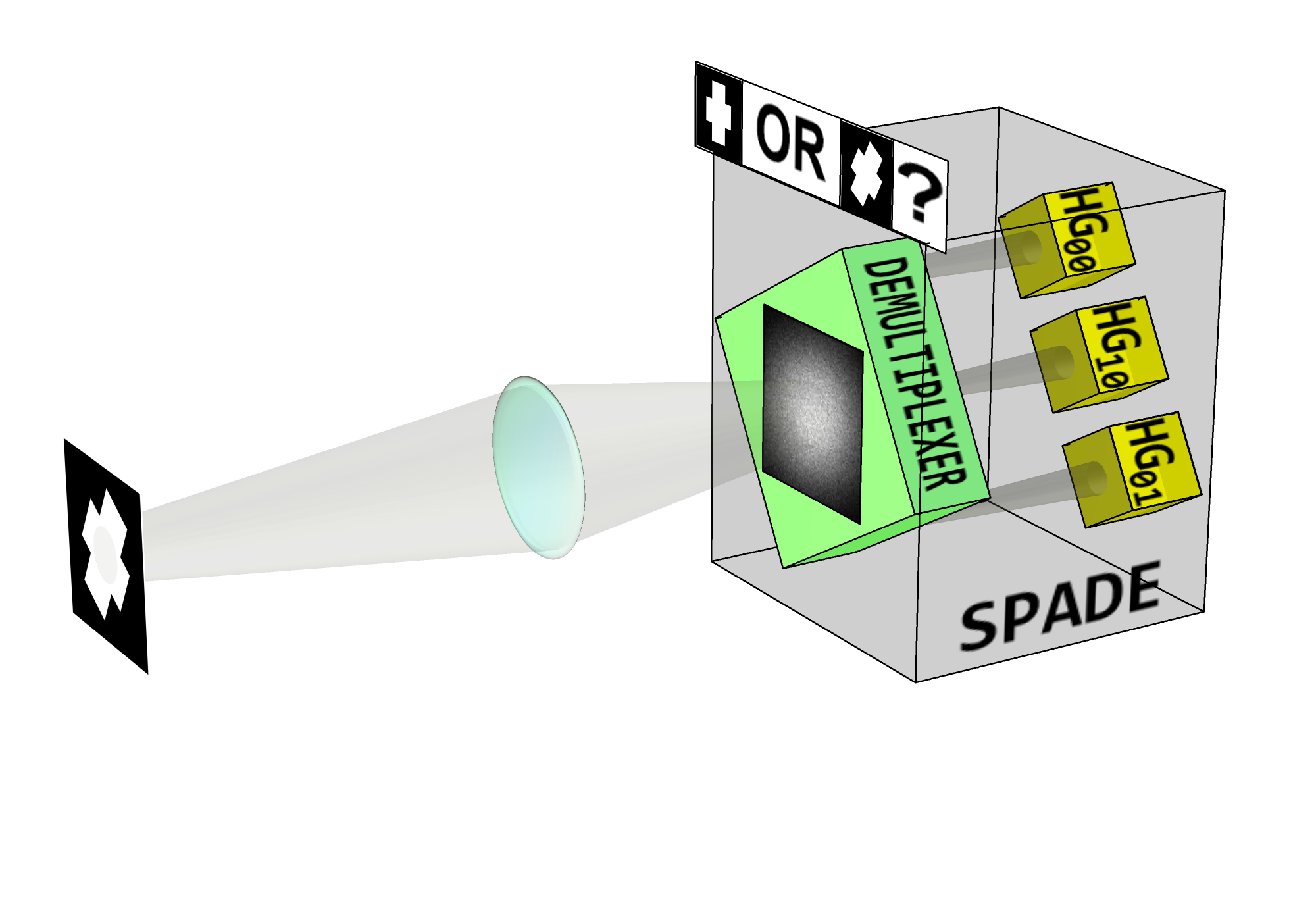}
    \caption{Schematic representation of the TRISPADE scheme for optimal identification of arbitrary incoherent sources in the subdiffraction regime.
    In the depicted example, the setup aims to identify the source between two hypotheses: a vertical cross, or a slightly tilted version of it.
    As discussed in Sec.~\ref{sec:subd}, the optimal local measurement is the projection onto a TRISPADE basis rotated of an hypotheses-dependent angle $\theta_0$.
    This can be achieved by simply rotating the demultiplexer.}
    \label{fig:setup}
\end{figure}

After choosing a measurement scheme, represented by a POVM $\Pi_k$ and the outcome probability $P_{i;k}=\Tr[\Pi_k\rho_i]$ conditioned on hypothesis $\rho_i$, we can calculate the asymptotic behavior of the probability of error $P_E=\e^{-N\xi}$ of the best discrimination strategy conceivable with $\Pi_k$ after a large number of repetitions $N$ of the measurement through the Chernoff exponent~\cite{Chernoff1952}
\begin{equation}
	\xi=-\ln\min_{s\in[0,1]}\sum_k P_{1;k}^sP_{2;k}^{1-s},
\end{equation}
where the sum runs over all the outcomes of the POVM.
We want to evaluate the Chernoff exponent $\xi$ associated with a measurement of the first $6$ relevant $HG$ modes.
Due to the common factorization of the density matrices $\rho_i=\rho_i^\alpha\otimes\rho_i^\beta$ and due to the property, shown in Eq.~\eqref{eq:IndistinguibleAlpha}, that the covariance matrices $\gamma_i^\alpha$ have only a nonzero eigenvalue (which is irrelevant in the calculation of the quantum Chernoff exponent), we will focus our analysis to a POVM projecting each photon onto the rotated SPADE basis $\{\ket{HG_{00}},\ket*{HG_{10}^{\theta_0}},\ket*{HG_{01}^{\theta_0}}\}$, also dubbed ``TRISPADE"~\cite{Grace2022}, 
where
\begin{gather}
	\ket*{HG^{\theta_0}_{1,0}}=\cos\theta_0\ket{HG_{1,0}}-\sin\theta_0\ket{HG_{0,1}},\\
    \ket*{HG^{\theta_0}_{0,1}}=\sin\theta_0\ket{HG_{1,0}}+\cos\theta_0\ket{HG_{0,1}},
\end{gather}
are $HG$ modes rotated of an angle $0\leqslant\theta_0\leqslant \frac{\pi}{2}$ (see \figurename~\ref{fig:setup}).
In Appendix~\ref{app:CB} we evaluate the probabilities $P_{i;N_0,j}$ of the relevant outcomes of the measurement up to order $\chi^2$ conditioned to hypothesis $j=1,2$, which can be identified by the number of photons $N_0$ detected in mode $HG_{0,0}$, and whether a photon is also detected in mode $HG_{1,0}^{\theta_0}$ ($i=1$), in mode $HG_{0,1}^{\theta_0}$ ($i=2$), or none of them ($i=0$), namely
\begin{gather}\begin{split} 
    P_{j;N_0,0}&= C_{N_0}\left(1-\chi^2\frac{N_0+I_0^2}{1+I_0}(V_{jx}+V_{jy})\right)+\O(\chi^4),\\
    P_{j;N_0,1}&= C_{N_0}\chi^2I_0\mathcal{V}_{jx}+\O(\chi^4),\\
    P_{j;N_0,2}&= C_{N_0}\chi^2I_0\mathcal{V}_{jy}+\O(\chi^4),
    \label{eq:Probs}
\end{split}\end{gather}
with $C_{N_0}=I_0^{N_0}(1+I_0)^{-(N_0+1)}$, and where
\begin{gather}\begin{split}
\mathcal{V}_{ix}&=\cos^2(\Delta_i)V_{ix}+\sin^2(\Delta_i)V_{iy} \, , \\
\mathcal{V}_{iy}&=\sin^2(\Delta_i)V_{ix}+\cos^2(\Delta_i)V_{iy},
\end{split}\label{eq:RotatedVar}\end{gather}
are the variances of the intensity distributions in the rotated reference frame, with $\Delta_i=\theta_i-\theta_0$, $i=1,2$.
We show in Appendix~\ref{app:CB} that the Chernoff exponent reads, neglecting terms of order $\O(\chi^4)$,
\begin{multline}
\xi(\theta_0)=\chi^2I_0\max_{s\in[0,1]}\Big[s(V_{1x}+V_{1y})+(1-s)(V_{2x}+V_{2y})\\
    -\mathcal{V}_{1x}^s\mathcal{V}_{2x}^{1-s}-\mathcal{V}_{1y}^s \mathcal{V}_{2y}^{1-s}\Big],
    \label{eq:CBGeneric}
\end{multline}
The optimal TRISPADE Chernoff exponent is thus given by the further maximization $\xi=\max_{\theta_0} \xi(\theta_0)$, i.e.
\begin{multline}
\xi=\chi^2I_0\max_{s\in[0,1]}\Big[s(V_{1x}+V_{1y})+(1-s)(V_{2x}+V_{2y})\\
   -\min_{\theta_0\in[0,\frac{\pi}{2}]}(\mathcal{V}_{1x}^s\mathcal{V}_{2x}^{1-s}+\mathcal{V}_{1y}^s \mathcal{V}_{2y}^{1-s})\Big].
   \label{eq:CBopt}
\end{multline}
We immediately notice that $\xi_Q$ in Eq.~\eqref{eq:QCB} and $\xi$ in Eq.~\eqref{eq:CBopt} are not generally equal.
In the rest of this section, we will give a closer inspection to some specific cases, identifying the scenarios when TRISPADE is optimal.
Nevertheless, the maximization in Eq.~\eqref{eq:CBopt} can always be solved numerically to find the best TRISPADE basis (identified by $\theta_0$) and to optimize the result of the identification experiment.

\subsection{Special identification scenarios}
\subsubsection{One point source}\label{sec:1Point}

If one of the two sources is a point-like source, SPADE becomes an optimal measurement.
Indeed, if we set $V_{1x}=V_{1y}=0$, while we rename $V_{2x}=V_x$, $V_{2y}=V_y$, the nonlinear terms of $\xi_Q$ in Eq.~\eqref{eq:QCB} and $\xi$ in Eq.~\eqref{eq:CBopt} disappear, so that
\begin{equation}
	\xi_Q=I_0\chi^2(V_x+V_y)=\xi.
\end{equation}
The optimality of SPADE for faint point-source discrimination was already demonstrated~\cite{Lu2018, Huang2021, Grace2022}.
However, it is interesting to see that the optimality is preserved for arbitrary intensities.

\subsubsection{Commuting covariance matrices}
\label{sec:commuting}

For commuting covariances matrices $\gamma_i^\beta$, TRISPADE is always optimal, provided one rotates the interferometer in order to align it with the common eigenbasis of the two matrices.
Indeed, it is possible to impose the commutativity between the two covariance matrices either by fixing $\Delta\theta=0,\frac{\pi}{2}$ or by choosing one covariance matrix to be proportional to the identity, say $V_{1x}=V_{1y}=V_1$.
In the first case, setting $\Delta\theta=0$ the quantum Chernoff exponent reduces to
\begin{multline}
	\xi_Q \overset{\Delta\theta=0}{=}I_0\chi^2\max_{s\in [0,1]}\Big[s(V_{1x}+V_{1y})+(1-s)(V_{2x}+V_{2y})\\
    -( V_{1x}^sV_{2x}^{1-s}+V_{1y}^sV_{2y}^{1-s})\Big],
\end{multline}
with the case $\Delta\theta=\frac{\pi}{2}$ obtained substituting $V_{2x}\leftrightarrow V_{2y}$. 
For the case $V_{1x}=V_{1y}=V_1$, simple algebra shows that
\begin{multline}
    \xi_Q \overset{V_{1x}=V_{1y}=V_1}{=}I_0\chi^2\max_{s\in [0,1]}\Big[2sV_1+(1-s)(V_{2x}+V_{2y})\\
    -V_1^s(V_{2x}^{1-s}+V_{2y}^{1-s})\Big].
\end{multline}

We can easily show that TRISPADE is optimal in both cases by comparing $\xi$ in Eq.~\eqref{eq:CBopt} with the two previous expressions of $\xi_Q$.
For $\Delta\theta=0$, we can choose $\theta_0=\theta_1=\theta_2$ so that $\Delta_1=\Delta_2=0$, which corresponds to aligning the interferometer along the common eigenbasis of the two covariance matrices, yielding $\mathcal{V}_{ix}=V_{ix}$ and $\mathcal{V}_{iy}=V_{iy}$, and thus $\xi=\xi_Q$.
For $V_{1x}=V_{1y}=V_1$, the measured variances of the first source in any rotated reference frame are $\mathcal{V}_{1x}=\mathcal{V}_{1y}=V_1$. 
If we align the interferometer to the second image, i.e. choosing $\theta_0=\theta_2$ so that $\mathcal{V}_{2x}=V_{2x}$ and $\mathcal{V}_{2y}=V_{2y}$, we have $\xi=\xi_Q$.
In other words, for commuting covariance matrices the prescription for quantum optimality consists in rotating the demultiplexer along the common principal variances of the two images.

\subsubsection{1D sources}\label{sec:1D}
Here we will consider $1$D images in a $2$D image plane.
Although it is a mathematical idealization, this case still provides interesting results that can be applied when the intensity distributions are highly concentrated along possibly different directions.
Optimality is always achieved when the two images are orthogonal and when they are identical but rotated.
Interestingly, we identify and discuss in detail some scenarios where SPADE is not optimal. 

We set $V_{1x}=V_{2x}=0$ and call $V_{1y}=V_1\neq 0$, $V_{2y}=V_2\neq 0$ without loss of generality, as the other cases produce identical expressions.
In this case, the quantum Chernoff coefficient becomes
\begin{equation}
	\xi_Q =I_0\chi^2\max_{s\in [0,1]}\Big[sV_1+(1-s)V_2
    -\cos^2\Delta\theta\ V_{1}^sV_{2}^{1-s}\Big].
    \label{eq:QCB1D}
\end{equation}
The maximization is analytically tractable but yield a convoluted expression. 
However, the maximization becomes trivial when $\Delta\theta=\pi/2$, i.e.~for orthogonal $1$D images
\begin{equation}
		\xi_Q \overset{\Delta\theta=\frac{\pi}{2}}{=}I_0\chi^2\max\{V_1,V_2\},
\end{equation}
and when $V_1=V_2=V$, i.e.~for identical but rotated $1$D images
\begin{equation}
			\xi_Q \overset{V_1=V_2=V}{=}I_0\chi^2V\sin^2\Delta\theta.
\end{equation}

To evaluate $\xi$, we first notice that the rotated variances in Eq.~\eqref{eq:RotatedVar} become $\mathcal{V}_{ix}=\sin^2(\Delta_i)V_i$, $\mathcal{V}_{iy}=\cos^2(\Delta_i)V_i$.
We thus have
\begin{multline}
	\xi=\chi^2I_0\max_{s\in[0,1]}\Big[sV_1+(1-s)V_2-V_1^sV_2^{1-s}\\    \times\min_{\theta_0\in[0,\pi]}\left((\sin^2\Delta_1)^s(\sin^2\Delta_2)^{1-s}+(\cos^2\Delta_1)^s(\cos^2\Delta_2)^{1-s}\right)\Big].
    \label{eq:CB1D}
\end{multline}
We want to find the condition for which $\xi=\xi_Q$.
For this purpose, we consider the inequality $(\sin^2\Delta_1)^s(\sin^2\Delta_2)^{1-s}+(\cos^2\Delta_1)^s(\cos^2\Delta_2)^{1-s}\geqslant\cos^2(\Delta\theta)$, with the saturation achieved for $\Delta\theta=0$ independently of $\theta_0$ and $s$, and for $\theta_0=\theta_1$ ($\theta_0=\theta_2$) if either $s=0$ ($s=1$), or $\Delta\theta=\pi/2$ (see Appendix~\ref{app:Ineq}).
\begin{figure}
    \centering
    \includegraphics[width=0.95\linewidth]{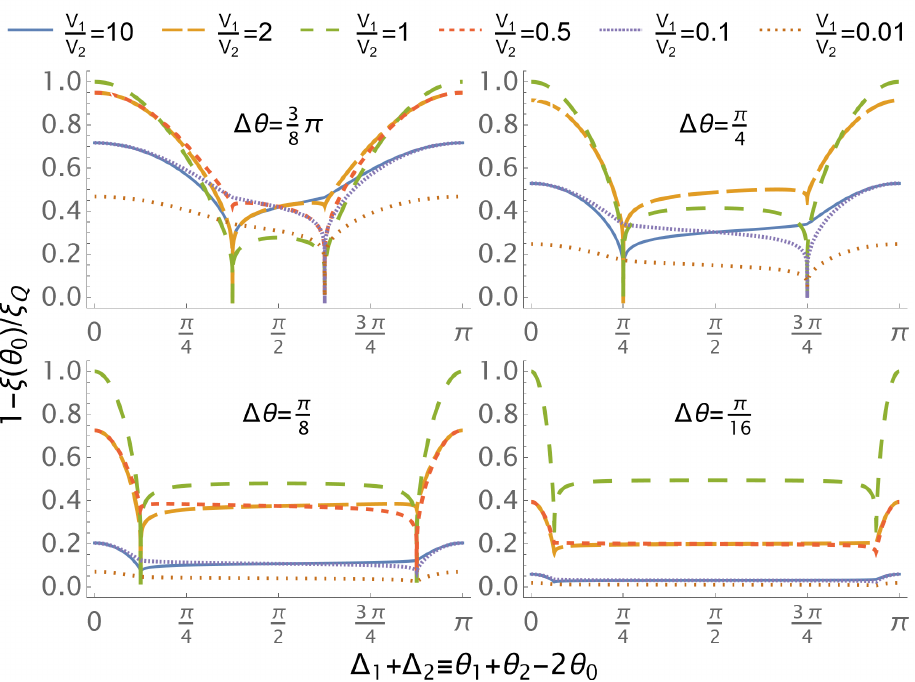}
    \caption{Plots of the normalized gap between the quantum Chernoff exponent $\xi_Q$ and the Chernoff exponent $\xi(\theta_0)$ associated with the rotated TRISPADE for the discrimination of 1D sources (see Section~\ref{sec:1D}), while varying $\Delta_1+\Delta_2$, that is linear in the SPADE rotation angle $\theta_0$: to $\Delta_1+\Delta_2=0$ corresponds $\theta_0=(\theta_1+\theta_2)/2$, while to $\Delta_1+\Delta_2=\pi$ corresponds $\theta_0=(\theta_1+\theta_2)/2-\pi/2$}
    \label{fig:1D}
\end{figure}
However, we notice that for $\Delta\theta=\pi/2$, the maximization over $s$ in Eq.~\eqref{eq:QCB1D} is still achieved for $s=0$ if $V_2\geqslant V_1$ or $s=1$ if $V_1\geqslant V_2$.

The previous analysis ultimately tells us that TRISPADE for $1$D images discrimination is optimal either with parallel $1$D images ($\Delta\theta=0$) without the need of optimizing $\theta_0$, for which
    \begin{equation}
	\xi=\xi_Q=I_0\chi^2\max_{s\in [0,1]}\Big[sV_1+(1-s)V_2
    -V_{1}^sV_{2}^{1-s}\Big],
    \end{equation}
or when the maximization over $s$ in Eq.~\eqref{eq:QCB1D} is achieved for $s=0$ or $s=1$, in which case the SPADE needs to be rotated at an angle $\theta_0=\theta_1$ or $\theta_0=\theta_2$, respectively. 
One can check with simple algebra that the maximization occurs for $s=0$ ($s=1$) when $\cos^2\Delta\theta\log r\geqslant r-1$, with $r=V_1/V_2$ ($r=V_2/V_1$).
This defines a $\Delta\theta$-dependent interval of values of $V_1/V_2\in[r^*(\Delta\theta),r^*(\Delta\theta)^{-1}]$, with $r^*(0)=1$ and $r^*(\pi/2)=0$, for which SPADE is optimal provided that $\theta_0$ is rotated along the image with smaller variance, with
\begin{equation}
\xi=\xi_Q=I_0\chi^2\sin^2\Delta\theta\max\{V_1,V_2\}.
\label{eq:1D}
\end{equation}
In Fig.~\ref{fig:1D} we verify this behavior numerically.
For wider angles, e.g.~$\Delta\theta=3\pi/8$, optimality is achieved either for $\theta_0=\theta_1$ or $\theta_0=\theta_2$ for a wide range of ratios of the variances $V_1/V_2$, while such a range shrinks for $\Delta\theta$ closer to 0.
Although the optimality is lost when one variance is much larger than the other, we can also notice that for increasing $r$ we asymptotically reach the optimal case of one point source identification discussed in Sec.~\ref{sec:1Point}.
Indeed, we can see from the plots of \figurename~\ref{fig:1D} that $\xi(\theta_0)$ slowly approaches $\xi_Q$ independently of $\theta_0$ as the ratio between the two variances gets values far from $1$. 
We can identify two more interesting limit cases: for orthogonal $1$D images, 
$\Delta\theta=\pi/2$, optimality is achieved for all values of $V_1$ and $V_2$ when aligning the interferometer to either of the images $\theta_0=\theta_1,\theta_2$; 
for identical but rotated $1$D images, i.e.~for $V_1=V_2$, optimality is achieved independently of $\Delta\theta$ when aligning the interferometer to either of the images.

Notice that only for $\Delta\theta=0,\pi/2$ the covariance matrices $\gamma_i^\beta$ commute. 
Surprisingly, we have found that optimality through SPADE is achieved also in specific cases with non-commuting covariance matrices.

\subsubsection{Rotated sources}\label{sec:rotated}
\begin{figure}
    \centering
    \includegraphics[width=.95\linewidth]{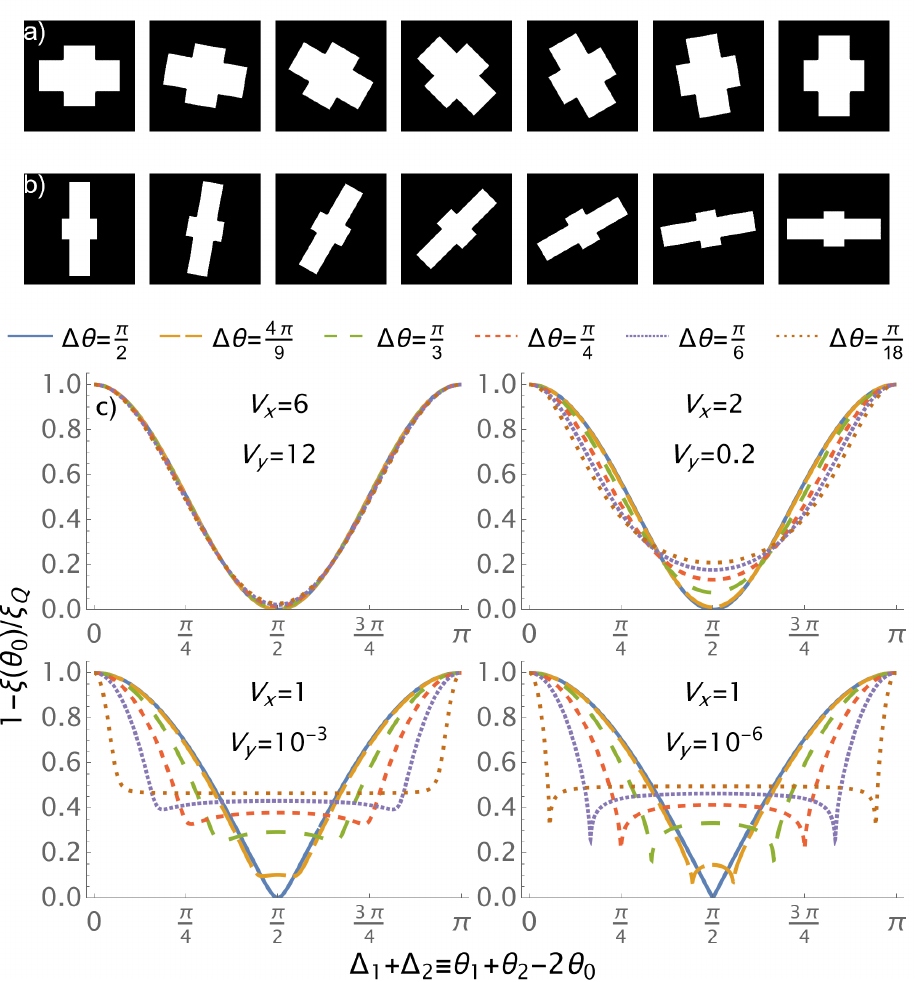}
    \caption{
    Analysis of the Chernoff bound for the identification of identical but rotated sources for a range of values of $\Delta\theta$ and of variances $V_x,V_y$.
    The variances have been chosen so that $\sqrt{V_x}-\sqrt{V_y}\simeq1$ in Eq.~\eqref{eq:QCBrotated}. 
    In particular the pairs $(V_x,V_y)=(6,12),\,(2,0.2)$ represent the variances of the images depicted on top in rows a) and b) respectively, rotated of different angles.
    c) Plots of the normalized gap between the quantum Chernoff exponent $\xi_Q$ and the Chernoff exponent $\xi(\theta_0)$ associated with the rotated TRISPADE (see Section~\ref{sec:rotated}), while varying $\Delta_1+\Delta_2$, ultimately controlled by the SPADE rotation angle $\theta_0$.
    }
    \label{fig:Rotated}
\end{figure}

Here, we examine the case where we are tasked to identify the orientation of the intensity distribution of a source. 
This can also be useful for checking the correct alignment of the interferometric setup with respect to a preferred orientation.
We see that optimality is retrieved when the two images are rotated of $90$ degrees.

We set $V_{1x}=V_{2x}=V_x\neq 0$ and $V_{1y}=V_{2y}=V_y\neq 0$, i.e.~the two images are identical up to the second moments, but they are also rotated of a relative angle $\Delta\theta$.
After simple algebra we obtain
\begin{multline}
	\xi_Q=\chi^2I_0\sin^2\Delta\theta\left[V_x+V_y-\min_{s\in[0,1]}(V_x^sV_y^{1-s}+V_y^sV_x^{1-s})\right]\\
    =\chi^2I_0\sin^2\Delta\theta\left(\sqrt{V_x}-\sqrt{V_y}\right)^2,
    \label{eq:QCBrotated}
\end{multline}
as the minimization is achieved for $s=1/2$.
The rotated variances become $\mathcal{V}_{ix}=\cos^2\Delta_i V_x+\sin^2\Delta_iV_y$, $\mathcal{V}_{iy}=\sin^2\Delta_i V_x+\cos^2\Delta_iV_y$, so that
\begin{equation}
	\xi=\chi^2I_0\Bigg[V_x+V_y-\min_{\substack{s\in[0,1]\\\theta_0\in[0,\frac{\pi}{2}]}}(\mathcal{V}_{1x}^s\mathcal{V}_{2x}^{1-s}+\mathcal{V}_{1y}^s \mathcal{V}_{2y}^{1-s})\Bigg] \, .
    \label{eq:CBRotated}
\end{equation}

Excluding the limiting cases where $\theta_1=\theta_2$ and $V_x=V_y$, for which the two sources become identical and thus $\xi_Q=0$, we can verify directly that the saturation $\xi=\xi_Q$ is achieved when $\theta_1=\theta_2\pm\pi/2$, i.e.~when the two images are rotated of 90 degrees, and choosing $\theta_0=\theta_1,\theta_2$.
Indeed this scenario is a particular case of commuting covariance matrices discussed in Sec.~\ref{sec:commuting}, and we have
\begin{equation}
	\xi\overset{\Delta\theta=\frac{\pi}{2}}{=}\xi_Q
=\chi^2 I_0 \left(\sqrt{V_x}-\sqrt{V_y}\right)^2.
\end{equation}
We can see the suboptimality of TRISPADE when this condition is not verified in the plots of Fig.~\ref{fig:Rotated}.
For images with sufficient spherical symmetry ($V_x\simeq V_y$), the gap $(\xi_Q-\xi)/\xi_Q\simeq0$ is small for the choice of $\theta_0=(\theta_1+\theta_2)/2\pm\pi/4$.
Although the quantum optimality is lost for $\Delta\theta\neq \pi/2$, the value of $\theta_0=(\theta_1+\theta_2)/2\pm\pi/4$ remains the best choice for every value of $\Delta\theta$ until the intensity distribution considered becomes highly "stretched".
In this case, the optimal $\theta_0$ depends on the values of $V_x$, $V_y$ and $\Delta\theta$, and it should be found numerically by solving Eq.~\eqref{eq:CBRotated}.
Noticeably, in \figurename~\ref{fig:Rotated} cusps appear at $\theta_0=\theta_1,\theta_2$ for $V_y\rightarrow0$, as the image tends to a $1$D representation, recovering the result in Eq.~\eqref{eq:1D}.

\section{Hypothesis testing for finite $N$}\label{sec:Bayes}

Both the quantum Chernoff exponent $\xi_Q$ and the Chernoff exponent $\xi$ associated with the TRISPADE measurement only determine the exponential decay rates of the probability of error $P_E$ of the optimal hypothesis testing in the asymptotic regime of a large number $N$ of frames.
The saturation of the quantum Chernoff bound $\xi_Q=\xi$ found under specific conditions discussed in the previous section thus only guarantees that a hypothesis test based on a TRISPADE local measurement can achieve such a decay rate for `sufficiently' large $N$.
In this section we will analyze the performance of the Bayesian hypothesis test in the sub-diffraction regime for finite $N$ and quantify when the asymptotic behavior is practically achieved.

\begin{figure}
    \centering
    \includegraphics[width=\linewidth]{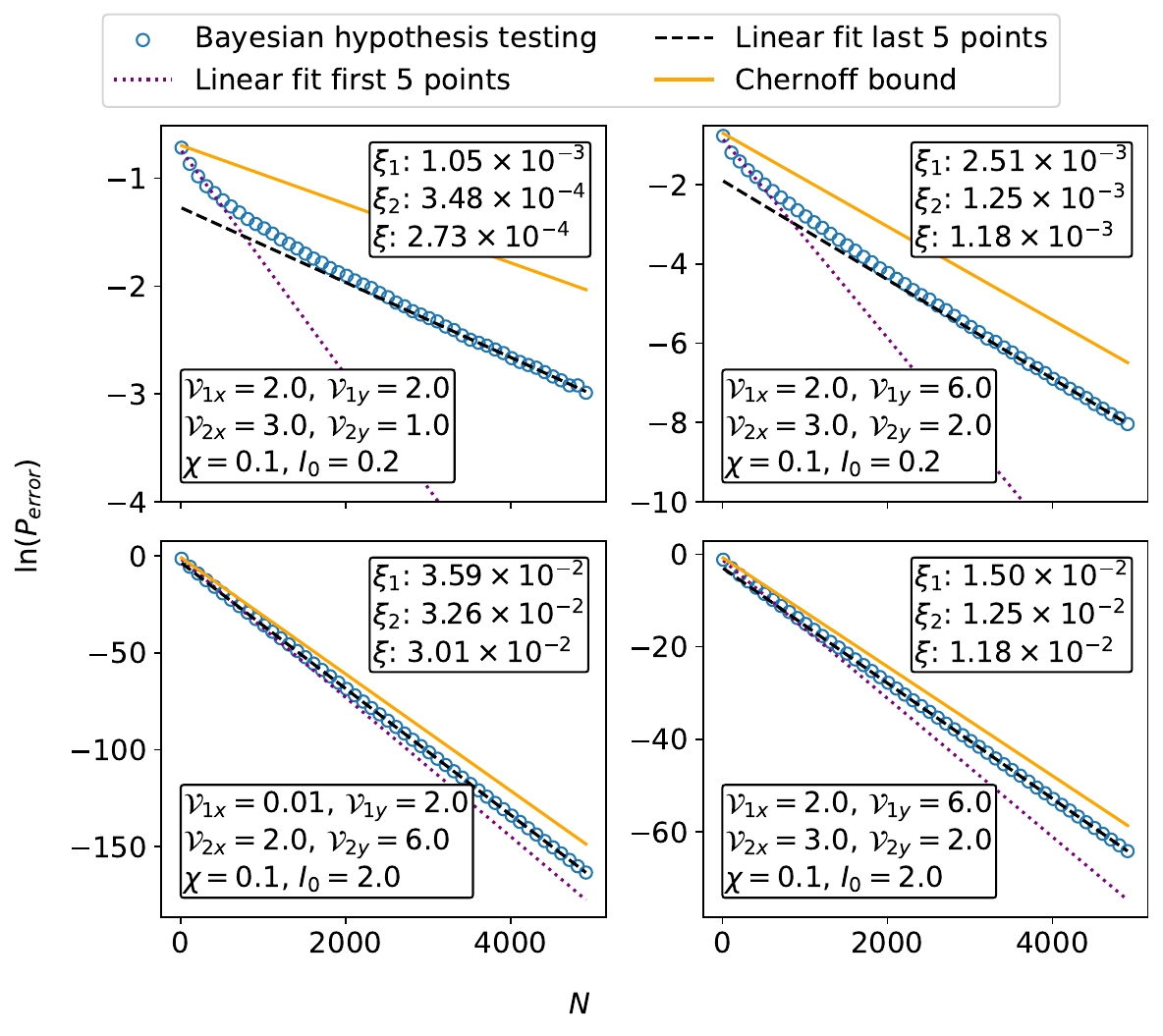}
    \caption{Plots of the simulated logarithm of the error probability $\ln P_e$ (blue circles) of the Bayesian test discussed in Sec.~\ref{sec:Bayes}, for different values of the variances, sub-diffraction parameter and average intensity. The pink dotted and the black dashed lines are the linear fit of the first ($10\leq N\leq 510$) and last ($4510\leq N\leq 5010$) five points respectively, whose slopes are reported as $\xi_1$ and $\xi_2$ in the upper right boxes. The orange solid line represents the optimal asymptotic scaling given by the Chernoff bound, whose offset has been arbitrarily chosen so that for $N=0$ the probability of error is $1/2$.}
    \label{fig:Bayes}
\end{figure}
For each frame, the outcome of the measurement is identified by the number of photons detected in the modes $HG_{00}$, $HG_{10}^{\theta_0}$ and $HG_{01}^{\theta_0}$, for which we can evaluate the probability $P_{j;N_0,i}$ to occur conditioned by either hypothesis $\rho_j$, $j=1,2$ shown in Eq.~\eqref{eq:Probs}.
By the Bayes' theorem, we can relate these conditioned probabilities to the posterior probability $P(\rho_j|N_0,i)=P_{j;N_0,i} P(\rho_j)/P(N_0,i)$ that hypothesis $\rho_j$ is true given the observed outcome, where $P(\rho_j)$ and $P(N_0,i)$ are the prior probability of the hypothesis $\rho_j$ and the marginal probability of the outcomes respectively.
Assuming complete lack of knowledge on which hypothesis is true, so that $P(\rho_j)=1/2$, we can then define the likelihood ratio $\lambda_{N_0,i}=P_{1;N_0,i}/P_{2;N_0,i}=P(\rho_1|N_0,i)/P(\rho_2|N_0,i)$ of the observed outcome, and we notice that $\lambda_{N_0,i}>1$ corresponds to a higher probability that hypothesis $\rho_1$ is true, given the observed outcome.
If we independently repeat the measurement $N$ times, with similar steps we can calculate the total likelihood ratio $\Lambda=\prod_{k=1}^N\lambda_{N_{0,k},i_k}$, that we can use to formulate a test, accepting hypothesis $\rho_1$ if $\Lambda>1$, $\rho_2$ if $\Lambda<1$, or flipping a coin if $\Lambda=1$.
The Bayesian test also provides a confidence level $C$ in the decision, quantified by the posterior probability itself, with $C=\max\{\Lambda/(1+\Lambda),1/(1+\Lambda)\}$.
Finally, we can retrieve the success probability $P_s$ of the test as the expected value of the confidence level $C$
\begin{equation}
    P_s=\frac{P(\Lambda>1|\rho_1)+P(\Lambda<1|\rho_2)+P(\Lambda=1)}{2}.
    \label{eq:Success}
\end{equation}
In principle, the success probability $P_s$ can be evaluated exactly.
However, as the number of frames $N$ increases, the conditional probabilities $P(\Lambda|\rho_j)$ become increasingly hard-to-compute multinomial distributions.

We then evaluate the probability of error $P_e=1-P_s$ numerically, performing a Monte Carlo simulation for up to $N\simeq5000$ repetitions of the measurement.
In particular, we developed a simplified Bayesian hypothesis test that does not require photon-resolving detectors to measure $N_0$, but instead, it only requires to count three types of detection events: a photon in the mode $HG_{10}$ (type-1 event), a photon in the mode $HG_{01}$ (type-2 event), or only photons in mode $HG_{00}$ (type-0 event). 
The details for the justification and implementation of this test can be found in Appendix~\ref{app:Bayes}, while the  results of the numerical simulations are shown in \figurename~\ref{fig:Bayes}.
We see that, for all four examples simulated, the rate of decay of the error probability, given by the slope $\xi_{sim}(N)$ of $-\ln(P_e)$ as a function of $N$ reaches values comparable to the theoretical Chernoff exponent $\abs{\xi_{sim}(N)-\xi}\ll\O(\chi^2)$ already for $N\leqslant 5000$.
Moreover, $\ln(P_e)$ appears to be a convex function of $N$ that approaches the optimal asymptotic scaling $\xi$ from above, i.e. $\xi_{sim}(N)\geqslant\xi$ for finite $N$.
Finally, we notice that, as expected from the theoretical results, larger values of $I_0$ greatly improve the efficiency of the test, with the test achieving in the simulated examples a $2\sigma\simeq95\%$ confidence to be correct ($\ln P_e\simeq-3$) already for $N\lesssim 100$ when receiving in average $2$ photons per frame from the thermal source ($I_0=2$), while $N\lesssim 5000$ are required for $0.2$ photons per frame ($I_0=0.2$).

\section{Conclusions}

We have presented a general theoretical framework for modeling arbitrary-intensity incoherent optical sources subject to diffraction. The formalism, based on the Glauber–Sudarshan representations of Gaussian states, applies broadly beyond the discrimination task considered here, offering a versatile tool for quantum optical analyses of incoherent light.
Specializing our results to Gaussian point-spread functions in the subdiffraction regime, we have shown that SPADE measurements are not universally optimal for discrimination: although the Hermite-Gauss basis is a natural choice for the description of the incoherent light in the subdiffraction regime for arbitrary intensities, SPADE in such a basis achieves the quantum Chernoff bound only when the intensity distributions under the two hypotheses satisfy specific compatibility conditions, such as commutative covariance matrices. 
Nevertheless, our analysis identifies the optimal SPADE strategy, assuming Gaussian point-spread functions, as the projection onto a hypotheses-dependent rotation of the Hermite-Gauss modes.
Finally, we developed a simplified Bayesian hypothesis test for the identification task at hand solely based on counting the number of frames where a photon is detected in the rotated mode $HG^\theta_{10}$, in $HG^\theta_{01}$, or neither of them, and therefore does not require photon-resolving detectors.
We numerically simulated the error probability of such a test, an we showed that its decay rate practically achieves the classical Chernoff exponent for a small number of frames $N\leqslant5000$, with the value of the minimum $N$ reducing to a few hundreds for intense sources.

These results fill an important gap in the understanding of subdiffraction identification tasks for bright and generic sources, simultaneously highlighting the limits of SPADE optimality and the ingredients required to approach quantum-limited performance.
In the cases where SPADE fails to be optimal, it remains an open question to determine the optimal measurement that saturates the quantum Chernoff bound. The implementation of such a measurement may need collective operations unlike SPADE followed by photon detection, which is a local measurement with a semi-classical description.
A known optimal global measurement saturating the Chernoff bound is the Helstrom measurement on all the $N$ copies of the state~\cite{Helstrom1969}, but its practical implementation is usually difficult.
However, interesting developments in this direction have been made in the past years. 
For example, applications of quantum principle component analysis~\cite{Lloyd2014} has been proved to achieve enhanced optical imaging via quantum computation employing non-local measurements~\cite{Khabiboulline2019, Mokeev2025}.

If one decides to forgo the quantum optimality in favor of the practical realization of the best local strategy found in our results, an important future development is to take into consideration experimental nuisances such as crosstalk, losses, dark counts and misalignment.
The first three effects can be easily modeled by introducing a linear transformation of the outcome probabilities, such as a mixing stochastic matrix to simulate crosstalk.
Misalignment can be considered by reintroducing the first moments of the intensity distributions of the two sources appearing in the covariance matrix of the thermal state.
Future work may also extend this approach to partially coherent sources, hence going beyond the Gaussian nature of thermal states, or to adaptive or machine-learning-assisted non-local measurement strategies. 
Finally, this work and the model developed within it can serve as a stepping stone towards more complex but more practical identification tasks, such as multi-hypotheses or composite hypothesis testing~\cite{Berta2021, Lami2025}, in which the objective is to identify the physical quantum state out of two families of possible states, useful in classification tasks.

\acknowledgements{
    We acknowledge: 
    the Italian Space Agency (ASI, Agenzia Spaziale Italiana) through the project Subdiffraction Quantum Imaging (SQI) n. 2023-13-HH.0;
    the European Union -- Next Generation EU: NRRP Initiative, Mission 4, Component 2, Investment 1.3 -- D.D. MUR n. 341 del 15.03.2022 -- Next Generation EU (PE0000023 ``National Quantum Science and Technology Institute'');
    and Next Generation EU, Missione 4 Componente 1, PRIN 2022, project title ``QUEXO'', CUP: D53D23002850006.}

\newpage
\onecolumngrid
\appendix

\section{Unitary transformation in the Glauber-Sudarshan representation}\label{app:Model}

Given a $M$-mode quantum state
\begin{equation}
	\rho=\int\prod_{i=1}^M\dd\alpha^2_i\ P(\{\alpha_i\}) \bigotimes_{i=1}^M \ketbra{\alpha_i}
    \label{eq:GenericThermalApp}
\end{equation}
undergoing a unitary evolution $\mathcal U(a^\dag_i)=\sum_{j=1}^M U_{ij}^*b_j^\dag$, we want to evaluate $\rho'=\mathcal{U}(\rho)$.
We first notice that
\begin{equation}
	\mathcal U\left(\ket{\alpha_i}\right)=\exp(\alpha_i\sum_{j=1}^M U_{ij}^*b_j^\dag-\alpha_i^*\sum_{j=1}^MU_{ij}b_j)\ket{0},
\end{equation}
hence
\begin{equation}
	\mathcal U\left(\bigotimes_{i=1}^M\ket{\alpha_i}\right)=\exp(\sum_{i,j=1}^M \alpha_iU_{ij}^*b_j^\dag-\alpha_i^*U_{ij}b_j)\ket{0}=\exp(\sum_{j=1}^M\beta_j b_j^\dag-\beta_j^*b_j)\ket{0}=\bigotimes_{j=1}^M\ket{\beta_j},
\end{equation}
with $\beta_j=\sum_{i=1}^MU_{ji}^\dag \alpha_i$.
Recalling the unitarity of $\mathcal U$, we can thus effortlessly change integration variable $\dd\alpha^2_i\rightarrow\dd\beta_i^2$ in Eq.~\eqref{eq:GenericThermalApp}, obtaining
\begin{equation}
	\mathcal{U}(\rho)=\int\prod_{i=1}^M\dd\beta^2_i\ P\left(\left\{(U\beta)_i\right\}\right) \bigotimes_{i=1}^M \ketbra{\beta_i}{\beta_i}.
\end{equation}
If $\rho$ is a thermal state with covariance matrix $\Gamma$, so that $P(\{\alpha_i\})=\frac{1}{\sqrt{\det\pi\Gamma}}\exp(-\alpha^\dag\Gamma^{-1}\alpha)$, the $P$ function of the transformed state $\rho'$ reads $P\left(\left\{(U\beta)_i\right\}\right)=\frac{1}{\sqrt{\det\pi\Gamma'}}\exp(-\beta^\dag\Gamma'^{-1}\beta)$, with $\Gamma'=U^\dag\Gamma U$.

\section{Evaluation of the qunatum Chernoff exponent}\label{app:QCB}

The quantum Chernoff exponent for two Gaussian states with first momenta equal to zero is reported in the review by Weedbrook et al. as~\cite{Weedbrook2012}
\begin{equation}
     \xi_Q=-\ln\min_{s\in [0,1]}\left(2^M \sqrt{\frac{\det(g'_s(V_1^\oplus)g'_{1-s}(V_2^\oplus))}{\det(S_1\lambda'_s(V_1^\oplus)S_1^T+S_2\lambda'_{1-s}(V_2^\oplus)S_2^T)}}\right),
     \label{eq:QCBApp}
\end{equation}
in terms of the Williamson decompositions of the $2M\times2M$ covariance matrices $V_{1/2}=S_{1/2}V_{1/2}^\oplus S_{1/2}^T$ of the Wigner function of $\rho_{1/2}$, with $g'_s(x)=\frac{2^s}{(x+1)^s-(x-1)^s}$, $\lambda'_s(x)=\frac{(x+1)^s+(x-1)^s}{(x+1)^s-(x-1)^s}$.
To retrieve the expression in Eq.~\eqref{eq:GaussianQCB}, we first recall the relationship $V_i=2\Gamma_i+1$ between the $2M\times2M$ covariance matrices in the Wigner and Glauber representations.
Then, due to the block structure of $\Gamma_i=\gamma_i\otimes I_2$ and thus of $V_i$, we have $S_i=U_i\otimes I_2$, with $U_i$ matrix of the eigenvectors of $\gamma_i$, and the symplectic eigenvalues in $V_i^\oplus$ coincide with the eigenvalues of $2\gamma_i+1$, so that the matrices in the numerator and denominator of Eq.~\eqref{eq:QCBApp} retain the same block-diagonal structure and can be written in terms of $\gamma_i$.
If we define $g_s(x)=g'_s(2x+1)$ and $\lambda_s(x)=\lambda'_s(2x+1)$ and noticing that $\det(A\otimes I_2)=\det(A)^2$, we retrieve Eq.~\eqref{eq:GaussianQCB}.

\section{Retrieving the quantum Chernoff coefficient for faint sources}\label{app:Faint}

Since the covariance matrix $\gamma_k$, $k=1,2$, which enters as argument of $g_s$ and $\lambda_s$ in Eq.~\eqref{eq:GaussianQCB}, is proportional to the overall intensity $I_0$ (assuming to keep the relative pixel intensity $I_{i,j}/I_0$ constant), we can employ the approximations 
\begin{multline}
g_s(x)=\frac{1}{1+sx-x^s+\O(x^2)}=\sum_{n=0}^\infty(-sx+x^s)^n+\O(x^2)=1-sx+x^s+\sum_{n=2}^{1/s}x^{ns}+\O(x^{1+s})\\
=-sx+\sum_{n=0}^{1/s} x^{ns}+\O(x^{1+s})=(1-sx)\sum_{n=0}^{1/s} x^{ns}+\O(x^{1+s})=\frac{1-sx}{1-x^s}+\O(x^{1+s})
\end{multline}
\begin{multline}
\lambda_s(x)=\frac{1+sx+x^s+\O(x^2)}{1+sx-x^s+\O(x^2)}= (1+sx+x^s+\O(x^2))(-sx+\sum_{n=0}^{1/s} x^{ns}+\O(x^{1+s}))\\
=1+x^s+\sum_{n=1}^{1/s}x^{ns}+\sum_{n=1}^{1/s}x^{(n+1)s}+\O(x^{1+s})=1+2\sum_{n=1}^{1/s}x^{ns}+\O(x^{1+s})
\end{multline}
so that, if we shorten $\O(\min(I_0^{1+s},I_0^{2-s}))\rightarrow o$, noticing that the order of the terms in multiplying $x$ and $y$ is important, we get
\begin{multline}
(\lambda_s(x)+\lambda_{1-s}(y))^{-1}=\frac{1}{2}(1+\sum_{n=1}^{1/s}x^{ns}+\sum_{n=1}^{1/(1-s)}y^{n(1-s)}+o)^{-1}=\frac{1}{2}(-1+\sum_{n=0}^{1/s}x^{ns}+\sum_{n=0}^{1/(1-s)}y^{n(1-s)}+o)^{-1}\\
=\frac{1}{2}(-1+(1-x^s)^{-1}+(1-y^{1-s})^{-1}+o)^{-1}=\frac{1}{2}(1-x^s)(1-x^sy^{1-s})^{-1}(1-y^{1-s})+o
\end{multline}
and finally
\begin{equation}
g_s(x)(\lambda_s(x)+\lambda_{1-s}(y))^{-1}g_{1-s}(y)=\frac{1}{2}(1-sx)(1+x^sy^{1-s})(1-(1-s)y)+o
\end{equation}
We can thus evaluate
\begin{align}
\xi_Q&\overset{I_0\ll 1}{=}-\ln\min_{s\in [0,1]}\left[\det(1-s\gamma_1)\det(1-(1-s)\gamma_2)\det(1+\gamma_1^s\gamma_2^{1-s})\right]\notag\\
&\quad=\max_{s\in [0,1]}\left[s\tr[\gamma_1]+(1-s)\tr[\gamma_2]-\tr[\gamma_1^s\gamma_2^{1-s}]\right]\notag\\
&\quad=(I_0-\min_{s\in[0,1]}\tr[\gamma_1^s\gamma_2^{1-s}]),
\end{align}
as shown in Eq.~\eqref{eq:QCBWeak}.

\section{Obtaining the simplification useful to evaluate the quantum Chernoff exponent in the subdiffraction regime}
\label{app:Subdiff}

Applying the results of perturbation theory to evaluate the spectral decomposition of the submatrix 
\begin{equation}
    \gamma_i^\alpha=I_0
    \begin{pmatrix}
	    1-\chi^2(m_i^{2,0}+m_i^{0,2}) & \frac{\chi^2}{\sqrt{2}}m_i^{2,0} & \chi^2 m_i^{1,1} & \frac{\chi^2}{\sqrt{2}}m_i^{0,2}\\
        \frac{\chi^2}{\sqrt{2}}m_i^{2,0} & 0 & 0 & 0\\
        \chi^2 m_i^{1,1} & 0 & 0 & 0\\
        \frac{\chi^2}{\sqrt{2}}m_i^{0,2} & 0 & 0 & 0
	\end{pmatrix}=I_0
    \begin{pmatrix}
        \alpha_i & \chi^2\eta_i\\
        \chi^2\eta_i^T & 0_3
    \end{pmatrix},
\end{equation}
where $\alpha_i=(1-\chi^2(m_i^{2,0}+m_i^{0,2}))$, $\eta_i=(\frac{1}{\sqrt{2}}m_i^{2,0},m_i^{1,1},\frac{1}{\sqrt{2}}m_i^{0,2})$ and $0_3$ is the $3\times 3$ zero matrix, we easily see that it has only one generally nonzero eigenvalue, i.e. $I_0\alpha_i$, and associated eigenvector $v_i=(1, \chi^2\eta_i)^T$, up to order $\chi^2$.
The other three eigenvectors $u_{k,i}=(-\chi^2(\eta_i)_k,\delta_{1,k},\delta_{2,k},\delta_{3,k})^T$ for $k=1,2,3$, where $\delta$ denotes the Kronecker delta, are associated to the degenerate eigenvalue $0$.
We can call $U^\alpha_i=(v_i, u_{1,i}, u_{2,i}, u_{3,i})$ the matrix of such eigenvectors, so that $\gamma_i^\alpha=U^\alpha_i\diag(\alpha_i,0,0,0)(U_i^\alpha )^T$.
We can thus calculate with some algebra
\begin{equation}
\lambda_s(\gamma^\alpha_i)=U^\alpha_i\diag(\lambda_s(\alpha_i),0,0,0)(U_i^\alpha )^T=\begin{pmatrix}
        \frac{(1+I_0)^s+I_0^s}{(1+I_0)^s-I_0^s}-\chi^2\frac{2I_0^s(1+I_0)^{s-1}s(m_i^{2,0}+m_i^{0,2})}{((1+I_0)^s-I_0^s)^2} & 2\frac{I_0^s \chi^2\eta_i}{(1+I_0)^s-I_0^s}\\
        2\frac{I_0^s \chi^2\eta_i^T}{(1+I_0)^s-I_0^s} & 1_3
    \end{pmatrix},
\end{equation}
and applying the formula for the determinant of a block diagonal matrix
\begin{equation}
	\det\begin{pmatrix}
	    A & B\\
        C & D
	\end{pmatrix}=\det(D)\det(A-BD^{-1}C)
\end{equation}
to the matrix $\lambda_s(\gamma_1^\alpha)+\lambda_{1-s}(\gamma_2^\alpha)$, we notice that the piece $BD^{-1}C$ is of order $\chi^4$ and can be neglected.
We can thus write 
\begin{equation}
2^4\frac{\det(g_s(\lambda_1^\alpha))\det(g_{1-s}(\lambda_2^\alpha))}{\det(\lambda_s(\gamma_1^\alpha)+\lambda_{1-s}(\gamma_2^\alpha))}=2\frac{g_s(\alpha_1)g_{1-s}(\alpha_2)}{\lambda_s(\alpha_1)+\lambda_{1-s}(\alpha_2)}=\frac{1}{(1+\alpha_1)^s(1+\alpha_2)^{1-s}-\alpha_1^s\alpha_2^{1-s}}=1+\O(\chi^4)
\end{equation}

\section{Evaluating the Chernoff bound and exponent associated with a rotated SPADE}
\label{app:CB}

We would like now to evaluate the Classical Chernoff bound associated with the rotated SPADE discussed in Sec.~\ref{sec:CB}.
Since the covariance matrices are block-diagonal $\gamma_i=\gamma_i^\alpha\oplus\gamma_i^\beta$, the thermal state can be written as $\rho_i=\rho_i^\alpha\otimes\rho_i^\beta$.
The matrix $\gamma_i^\alpha$ has only one non-vanishing eigenvalue $\lambda_i^\alpha=I_0\alpha_i=I_0(1-\chi^2(m_i^{2,0}+m_i^{0,2}))$ up to order $\O(\chi^4)$ that coincides with its (1,1) element, whose associated eigenvector is $v_i=(1, \chi^2\eta_i)^T$, with $\eta_i=(\frac{1}{\sqrt{2}}m_i^{2,0},m_i^{1,1},\frac{1}{\sqrt{2}}m_i^{0,2})$, while the eigenvalues of $\gamma_i^\beta$ are both proportional to $\chi^2$, $\lambda_{i,x/y}^\beta=I_0\chi^2 V_{i,x/y}$.
We can thus write in the Fock basis
\begin{equation}
	\rho_i^\alpha=\sum_{m_0=0}^\infty \frac{(\lambda_i^\alpha)^{m_0}}{(1+\lambda_i^\alpha)^{m_0+1}}\ketbra{m_0}{m_0}_{\alpha,i},\qquad \ket{m_0}_{\alpha,i}=\frac{1}{\sqrt{m_0!}}(\hat{a}_{\alpha,i}^\dagger)^{m_0}\ket{0}.
\end{equation}
with $\hat{a}^\dagger_{\alpha,i}\ket{0}=(\hat{a}^\dagger_{0}+\chi^2\hat{a}^\dagger_{\eta_i})\ket{0}\equiv\ket{HG_{00}}+\chi^2\left((\eta_i)_1\ket{HG_{20}}+(\eta_i)_2\ket{HG_{11}}+(\eta_i)_3\ket{HG_{02}}\right)$, so that, calling $\ket{m_0}_0=\frac{1}{\sqrt{m_0!}}\hat{a}^{\dagger m_0}_{0}\ket{0}$ and $\ket{1}_{\eta_i}=\hat{a}^\dagger_{\eta_i}\ket{0}$, we can write
\begin{equation}
	\rho_i^\alpha=\frac{1}{1+\lambda_i^\alpha}\ketbra{0}{0} + \sum_{m_0=1}^\infty \frac{(\lambda_i^\alpha)^{m_0}}{(1+\lambda_i^\alpha)^{m_0+1}}\left(\ketbra{m_0}{m_0}_0+m_0\chi^2(\ket{m_0-1}_0\ket{1}_{\eta_i}\bra{m_0}_0+\ket{m_0}_0\bra{m_0-1}_0\bra{1}_{\eta_i})\right)+\O(\chi^4),
\end{equation}
and thus the probability of measuring $N_0$ photons in mode $\ket{HG_{00}}$ is
\begin{equation}
	P_{i;N_0,\alpha}=\frac{(\lambda_i^\alpha)^{N_0}}{(1+\lambda_i^\alpha)^{N_0+1}}+\O(\chi^4)=\frac{I_0^{N_0}}{(1+I_0)^{N_0+2}}\left(1+I_0+\chi^2(I_0-N_0)(V_{ix}+V_{iy})\right)+\O(\chi^4).
\end{equation}
Notice that ${}_{\eta_i}\!\bra{m}\rho_i^\alpha \ket{m}_{\eta_i}=\O(\chi^4)$, $\forall m$.
Regarding $\rho^\beta_i$, we can write
\begin{equation}
	\rho^\beta_i=\sum_{m_x=0}^\infty \frac{(\lambda_{i,x}^\beta)^{m_x}}{(1+\lambda_{i,x}^\beta)^{m_x+1}}\ketbra{m_x}{m_x}_{10,i} \otimes \sum_{m_y=0}^\infty\frac{(\lambda_{i,y}^\beta)^{m_y}}{(1+\lambda_{i,y}^\beta)^{m_y+1}}\ketbra{m_y}{m_y}_{01,i},
\end{equation}
where
\begin{gather}
    \ket{m_x}_{10.i}=\frac{1}{\sqrt{m_x!}}(\hat{a}^{\dagger}_{10,i})^{m_x}\ket{0},\qquad \hat{a}^{\dagger}_{10,i}\ket{0}=\ket{HG^{\theta_i}_{1,0}}=\cos\theta_i\ket{HG_{1,0}}-\sin\theta_i\ket{HG_{0,1}}\\
    \ket{m_y}_{01,i}=\frac{1}{\sqrt{m_y!}}(\hat{a}^{\dagger}_{01,i})^{m_y}\ket{0},\qquad \hat{a}^{\dagger}_{01,i}\ket{0}=\ket{HG^{\theta_i}_{0,1}}=\sin\theta_i\ket{HG_{1,0}}+\cos\theta_i\ket{HG_{0,1}}
\end{gather}
and
\begin{equation}
	\frac{(\lambda_{i,x/y}^\beta)^{m_{x/y}}}{(1+\lambda_{i,x/y}^\beta)^{m_{x/y}+1}}=\begin{cases}
    1-I_0\chi^2V_{ix/y}+\O(\chi^4),& m_{x/y}=0;\\
    I_0\chi^2V_{ix/y}+\O(\chi^4),& m_{x/y}=1;\\
    \O(\chi^4),&m_{x/y}\geqslant 2.
    \end{cases}
\end{equation}
so that
\begin{equation}
	\rho_i^\beta=(1-I_0\chi^2(V_{ix}+V_{iy}))\ketbra{0}{0}+ I_0\chi^2V_{ix}\hat{a}^\dagger_{10,i}\ketbra{0}{0}\hat{a}_{10,i}
    + I_0\chi^2V_{iy}\hat{a}^\dagger_{01,i}\ketbra{0}{0}\hat{a}_{01,i}+\O(\chi^4),
    \label{eq:RhoBeta}
\end{equation}
meaning that only the zero- and single-photon events have a non-negligible probability for small $\chi$.
Now we notice that a measurement along the rotated TRISPADE base $\ket*{HG_{1,0}^{\theta_0}},\ket*{HG_{1,0}^{\theta_0}}$ shown in~\ref{sec:CB} is equivalent to a counter-rotation of the reference frame of an angle $-\theta_0$, so that we can evaluate the probabilities of observing zero photons, one photon in $HG_{1,0}^{\theta_0}$, or one photon in $HG_{0,1}^{\theta_0}$ from $\rho_i^\beta$ in Eq.~\eqref{eq:RhoBeta} as
\begin{align}
    P_{i;0,0,\beta}&=1-I_0\chi^2(V_{ix}+V_{iy})+\O(\chi^4)\notag\\
    P_{i;1,0,\beta}&=I_0\chi^2(\cos^2(\theta_i-\theta_0)V_{ix}+\sin^2(\theta_i-\theta_0)V_{iy})+\O(\chi^4)\notag\\
    P_{i;0,1,\beta}&=I_0\chi^2(\sin^2(\theta_i-\theta_0)V_{ix}+\cos^2(\theta_i-\theta_0)V_{iy})+\O(\chi^4).
\end{align}

Finally, in order to evaluate the Chernoff exponent $\xi=-\ln\min_{s\in[0,1]}\sum_{k} P_{1;k}^s P_{2;k}^{1-s}$ were $k$ runs over all the observable events, we can group these outcome events in three cases:
\begin{enumerate}
    \item Observing $N_0$ $HG_{0,0}$ photons and no $HG_{0,1/1,0}^{\theta_0}$ photon, with probability 
    \begin{equation}	P_{i;N_0,0}=P_{i;N_0,\theta_0}P_{i;0,0,\beta}=\frac{I_0^{N_0}}{(1+I_0)^{N_0+2}}\left(1+I_0-\chi^2(I_0^2+N_0)(V_{ix}+V_{iy})\right)+\O(\chi^4)
    \end{equation}
    \item Observing $N_0$ $HG_{0,0}$, one $HG_{1,0}$ and no $HG_{0,1}$ photons, with probability 
    \begin{equation}   P_{i;N_0,1}=P_{i;N_0,\theta_0}P_{i;1,0,\beta}=\frac{I_0^{N_0}}{(1+I_0)^{N_0+2}}(\chi^2I_0(I_0+1)(\cos^2(\theta_i-\theta_0)V_{ix}+\sin^2(\theta_i-\theta_0)V_{iy}))
    \end{equation}
    \item Observing $N_0$ $HG_{0,0}$, no $HG_{1,0}$ and one $HG_{0,1}$ photons, with probability 
    \begin{equation}
	P_{i;N_0,2}=P_{i;N_0,\theta_0}P_{i;0,1,\beta}=\frac{I_0^{N_0}}{(1+I_0)^{N_0+2}}(\chi^2I_0(I_0+1)(\sin^2(\theta_i-\theta_0)V_{ix}+\cos^2(\theta_i-\theta_0)V_{iy}))
    \end{equation}

\end{enumerate}

We can finally evaluate the CB according to its definition
\begin{align}
    \xi = &-\ln\min_{s\in[0,1]}\left[\sum_{N_0=0}^\infty\sum_{j=0}^2 P_{1;N_0,j}^s P_{2;N_0,j}^{1-s}\right]=-\ln\min_{s\in[0,1]}\Bigg[\sum_{N_0=0}^\infty\frac{I_0^{N_0}}{(1+I_0)^{N_0+2}}\times\notag\\
    &\quad\times\Bigg(\left(1+I_0-\chi^2(I_0^2+N_0)(V_{1x}+V_{1y})\right)^s\left(1+I_0-\chi^2(I_0^2+N_0)(V_{2x}+V_{2y})^{1-s}\right)^{1-s}\notag\\
    &\quad+\chi^2 I_0(1+I_0)\Big((\cos^2(\theta_1-\theta_0)V_{1x}+\sin^2(\theta_1-\theta_0)V_{1y})^s(\cos^2(\theta_2-\theta_0)V_{2x}+\sin^2(\theta_2-\theta_0)V_{2y})^{1-s}\notag\\
    &\qquad\qquad\qquad\quad+(\sin^2(\theta_1-\theta_0)V_{1x}+\cos^2(\theta_1-\theta_0)V_{1y})^s(\sin^2(\theta_2-\theta_0)V_{2x}+\cos^2(\theta_2-\theta_0)V_{2y})^{1-s}\Big)\Bigg)\Bigg]\notag\\
    &=-\ln\min_{s\in[0,1]}\Bigg[\sum_{N_0=0}^\infty\frac{I_0^{N_0}}{(1+I_0)^{N_0+1}}\Bigg(1-\chi^2\frac{I_0^2+N_0}{1+I_0}\left(s(V_{1x}+V_{1y})+(1-s)(V_{2x}+V_{2y})\right)\notag\\
    &\quad+\chi^2I_0\Big((\cos^2(\theta_1-\theta_0)V_{1x}+\sin^2(\theta_1-\theta_0)V_{1y})^s(\cos^2(\theta_2-\theta_0)V_{2x}+\sin^2(\theta_2-\theta_0)V_{2y})^{1-s}\notag\\
    &\qquad\qquad+(\sin^2(\theta_1-\theta_0)V_{1x}+\cos^2(\theta_1-\theta_0)V_{1y})^s(\sin^2(\theta_2-\theta_0)V_{2x}+\cos^2(\theta_2-\theta_0)V_{2y})^{1-s}\Big)\Bigg)\Bigg]\notag\\
    &=-\ln\min_{s\in[0,1]}\Bigg[1-\chi^2I_0\Big(s(V_{1x}+V_{1y})+(1-s)(V_{2x}+V_{2y})\notag\\
    &\qquad\qquad-(\cos^2(\theta_1-\theta_0)V_{1x}+\sin^2(\theta_1-\theta_0)V_{1y})^s(\cos^2(\theta_2-\theta_0)V_{2x}+\sin^2(\theta_2-\theta_0)V_{2y})^{1-s}\notag\\
    &\qquad\qquad-(\sin^2(\theta_1-\theta_0)V_{1x}+\cos^2(\theta_1-\theta_0)V_{1y})^s(\sin^2(\theta_2-\theta_0)V_{2x}+\cos^2(\theta_2-\theta_0)V_{2y})^{1-s}\Big)\Bigg]\notag\\
    &=\chi^2I_0\max_{s\in[0,1]}\Bigg[s(V_{1x}+V_{1y})+(1-s)(V_{2x}+V_{2y})\notag\\
    &\qquad\qquad-(\cos^2(\theta_1-\theta_0)V_{1x}+\sin^2(\theta_1-\theta_0)V_{1y})^s(\cos^2(\theta_2-\theta_0)V_{2x}+\sin^2(\theta_2-\theta_0)V_{2y})^{1-s}\notag\\
    &\qquad\qquad-(\sin^2(\theta_1-\theta_0)V_{1x}+\cos^2(\theta_1-\theta_0)V_{1y})^s(\sin^2(\theta_2-\theta_0)V_{2x}+\cos^2(\theta_2-\theta_0)V_{2y})^{1-s}\Bigg],
    \label{eq:AppCBI0}
\end{align}
where we used
\begin{equation}
	\sum_{N_0=0}^\infty\frac{I_0^{N_0}}{(1+I_0)^{N_0+1}}=1,\qquad \sum_{N_0=0}^\infty\frac{I_0^{N_0}}{(1+I_0)^{N_0+1}}N_0=I_0,
\end{equation}
which coincides with the expression in Eq.~\eqref{eq:CBGeneric} once the substitutions in Eq.~\eqref{eq:RotatedVar} are employed.

\section{Proving an inequality useful in the analysis of the Chernoff exponent for 1D sources}\label{app:Ineq}

We want to show that
\begin{equation}
\label{eq:Ineqapp}
	(\sin^2\Delta_1)^s(\sin^2\Delta_2)^{1-s}+(\cos^2\Delta_1)^s(\cos^2\Delta_2)^{1-s}\geqslant\cos^2(\Delta\theta),
\end{equation}
and to find when saturation is achieved.
First, we notice that the expression on the left hand side of Eq.~\eqref{eq:Ineqapp} simplifies if $\Delta_1=0$ or $\Delta_2=0$, becoming, e.g. for $\Delta_1=0$
\begin{equation}
	\cos^2\Delta_2^{1-s}\geqslant\cos^2\Delta_2,
\end{equation}
which is evidently true and saturated either for $s=0$ or $\Delta_2=\frac{\pi}{2}$.

For the more general case, we invoke the inequality between arithmetic and geometric means (AM-GM) $x^sy^{1-s}\leqslant sx+(1-s)y$, so that
\begin{equation}
	(\sin^2\Delta_1)^{1-s}(\sin^2\Delta_2)^s+(\cos^2\Delta_1)^{1-s}(\cos^2\Delta_2)^{s}\leqslant (1-s)(\sin^2\Delta_1+\cos^2\Delta_1)+s(\sin^2\Delta_2+\cos^2\Delta_2)=1.
\end{equation}
Moreover, if we define the vectors $u=(\sin^s\Delta_1\sin^{1-s}\Delta_2,\cos^s\Delta_1\cos^{1-s}\Delta_2)$, $v=(\sin^{1-s}\Delta_1\sin^s\Delta_2,\cos^{1-s}\Delta_1\cos^s\Delta_2)$, by the Cauchy–Schwarz inequality we have
\begin{multline}
	\left((\sin^2\Delta_1)^s(\sin^2\Delta_2)^{1-s}+(\cos^2\Delta_1)^s(\cos^2\Delta_2)^{1-s}\right)\left((\sin^2\Delta_1)^{1-s}(\sin^2\Delta_2)s+(\cos^2\Delta_1)^{1-s}(\cos^2\Delta_2)^{s}\right)\geqslant\\
    (\sin\Delta_1\sin\Delta_2+\cos\Delta_1\cos\Delta_2)^2=\cos^2\Delta\theta,
\end{multline}
so that putting the two inequalities together, we obtain Eq.~\eqref{eq:Ineqapp}.
The AM-GM inequality is saturated iff $x=y$, which requires $\Delta_1=\Delta_2=\Delta$ for $0\leqslant\Delta_1,\Delta_2\leqslant\frac{\pi}{2}$. 
This condition makes the vectors $u,v=(\sin\Delta,\cos\Delta)$ equal, so that the Cauchy-Schwarz inequality is also saturated.

\section{Bayesian hypothesis testing}\label{app:Bayes}

In this appendix we will evaluate the likelihood ratios $\lambda_k$, for $k$ running over all possible outcomes of a TRISPADE measurement, and we will define and describe the Bayesian hypothesis test employed to obtain numerically the plots of \figurename~\ref{fig:Bayes}.

As evaluated in Appendix~\ref{app:CB} and discussed in Sec.~\ref{sec:CB}, the outcomes of each iteration of the TRISPADE measurement are identified by the number $N_0$ of photons detected in mode $HG_{00}$, and whether a photon is detected in mode $HG_{10}^{\theta_0}$ ($i=1$), in mode $HG_{01}^{\theta_0}$ ($i=2$), or neither of these modes $(i=0)$. 
For simplicity, we will call these events of type $1$, $2$ and $0$ respectively.
These events happen with probability $P_{j;N_0,i}$ defined in Eq.~\eqref{eq:Probs} given the hypothesis $\rho_j$.
We can easily evaluate the likelihood ratios of these events
\begin{gather}
    \begin{split}
        \lambda_{N_0,0}&=\frac{1-\chi^2\frac{N_0+I_0^2}{1+I_0}(V_{1x}+V_{1y})}{1-\chi^2\frac{N_0+I_0^2}{1+I_0}(V_{2x}+V_{2y})}+\O(\chi^4)=1-\chi^2\frac{N_0+I_0^2}{1+I_0}(V_{1x}+V_{1y}-V_{2x}-V_{2y})+\O(\chi^4),\\
        \lambda_{N_0,1}&=\frac{\mathcal{V}_{1x}} {\mathcal{V}_{2x}} + \O(\chi^4) \\
        \lambda_{N_0,2}&=\frac{\mathcal{V}_{1y}} {\mathcal{V}_{2y}}+ \O(\chi^4).
    \end{split}
\end{gather}
We notice that the likelihood ratios associated with events of type $1$ and $2$ depend neither on the sub-diffraction geometric factor $\chi$, nor on $N_0$, up to order $\chi^2$.

Let us now imagine to independently repeat the measurement $N$ times.
The overall likelihood ratio $\Lambda$ can be evaluated as the product of the likelihood ratios of the outcomes of each measurements.
Since $\lambda_{N_0,i}$ for $i=1,2$ does not depend on $N_0$, it is irrelevant to record the number $N_0$ of photons detected in $HG_{00}$ for events of type $1$ and $2$.
Let us then call $M_i$ the number of events of type $i$, for $i=0,1,2$, and we notice that $M_1$ and $M_2$ also coincide with the number of photons detected in total, over $N$ repetitions, in modes $HG_{10}^{\theta_0}$ and $HG_{01}^{\theta_0}$.
Let us call $N_{0}^{(k)}$, for $k=1,\dots M_0$, the number of photons detected in mode $HG_{00}$ at the $k$th event of type $0$.
The overall likelihood ratio $\Lambda$ can thus be evaluated as
\begin{align}
	\Lambda&=\left(\frac{\mathcal{V}_{1x}} {\mathcal{V}_{2x}}\right)^{M_1}\left(\frac{\mathcal{V}_{1y}} {\mathcal{V}_{2y}}\right)^{M_2}\prod_{k=1}^{M_0}\left(1-\chi^2\frac{N_0^{(k)}+I_0^2}{1+I_0}(V_{1x}+V_{1y}-V_{2x}-V_{2y})\right)+\O(\chi^4)\notag\\
    &=\left(\frac{\mathcal{V}_{1x}} {\mathcal{V}_{2x}}\right)^{M_1}\left(\frac{\mathcal{V}_{1y}} {\mathcal{V}_{2y}}\right)^{M_2}\left(1-\chi^2\frac{(\sum_{k=1}^{M_0}N_0^{(k)})+M_0I_0^2}{1+I_0}(V_{1x}+V_{1y}-V_{2x}-V_{2y})\right)+\O(\chi^4)\notag\\
    &=\left(\frac{\mathcal{V}_{1x}} {\mathcal{V}_{2x}}\right)^{M_1}\left(\frac{\mathcal{V}_{1y}} {\mathcal{V}_{2y}}\right)^{M_2}\left(1-\chi^2\frac{M_0\langle N_0\rangle+M_0I_0^2}{1+I_0}(V_{1x}+V_{1y}-V_{2x}-V_{2y})\right)+\O(\chi^4),
    \label{eq:LikelihoodRatio}
\end{align}
and we see that it only depends on the total number of photons detected in mode $HG_{10}$, $HG_{01}$, and in mode $HG_{00}$ during events of type $0$, where $\langle N_0\rangle=\sum_{k=1}^{M_0}N_0^{(k)}/M_0$ is its average. 
The likelihood ratio $\Lambda$ in Eq.~\eqref{eq:LikelihoodRatio} can already be employed to perform Bayesian hypothesis testing as described in Sec.~\ref{sec:Bayes} of the main text.
However, we can further simplify the test if we recall that events of type $0$ are much more likely than the ones of type $1$ and $2$.
Indeed, from the probabilities in Eq.~\eqref{eq:Probs}, we can identify the average number of photons $I_0$ with $I_0=\langle N_0\rangle+\O(\chi^2)$, and thus write the overall likelihood ratio as
\begin{multline}
	\Lambda\simeq\Lambda(M_0,M_1,M_2)=\left(\frac{\mathcal{V}_{1x}} {\mathcal{V}_{2x}}\right)^{M_1}\left(\frac{\mathcal{V}_{1y}} {\mathcal{V}_{2y}}\right)^{M_2}\left(1-\chi^2I_0M_0(V_{1x}+V_{1y}-V_{2x}-V_{2y})\right)+\O(\chi^4)\\
	=\left(\frac{\mathcal{V}_{1x}} {\mathcal{V}_{2x}}\right)^{M_1}\left(\frac{\mathcal{V}_{1y}} {\mathcal{V}_{2y}}\right)^{M_2}\left(\frac{1-\chi^2I_0(V_{1x}+V_{1y})}{1-\chi^2I_0(V_{2x}+V_{2y})}\right)^{M_0}+\O(\chi^4),
    \label{eq:LikelihoodRatioSimplified}
\end{multline}
where the last expression is explicitly written as a ratio of probabilities and therefore is better suited to be employed for the test.
Noticeably, it only depends on the number of observed events of each type ($M_0$, $M_1$ and $M_2$), and it coincides with the likelihood ratio associated with trinomial distributions with probabilities
\begin{gather}
    \begin{split}
        P_{j;0}&=\sum_{N_0=0}^\infty P_{j;N_0,0}=1-\chi^2I_0(V_{jx}+V_{jy})+\O(\chi^4)\\
        P_{j;1}&=\sum_{N_0=0}^\infty P_{j;N_0,1}=\chi^2I_0\mathcal{V}_{jx}+\O(\chi^4)\\
        P_{j;2}&=\sum_{N_0=0}^\infty P_{j;N_0,2}=\chi^2I_0\mathcal{V}_{jy}+\O(\chi^4),
    \end{split}
    \label{eq:ProbsTrinomial}
\end{gather}
where $P_{j;N_0,i}$ are once again defined in Eq.~\eqref{eq:Probs}.

For our simulation, we employed the likelihood ratio $\Lambda$ in Eq.~\eqref{eq:LikelihoodRatioSimplified} to formulate the hypothesis testing and to evaluate numerically the error probability $P_e=1-P_s$ shown in the plots of \figurename~\ref{fig:Bayes}, complementary of the success probability $P_s$ of the test in Eq.~\eqref{eq:Success}.
The simulation consisted in repeating twice $10^4$ trials of the test for increasing values of $N$ (from $N=10$ to $N=5010$ in steps of $100$), where data were generated by sampling from a trinomial distribution with probabilities $P_{j;i}$ in Eq,~\eqref{eq:ProbsTrinomial} with $i=0,1,2$, and $j=1$ in the first half of the trials, $j=2$ in the second half.
The probability of error has been recovered by the ratio of wrong guesses of the likelihood ratio test. 
Finally linear fits of the logarithm of the error probability versus the first $5$ values and last $5$ values of $N$ have been performed to retrieve the decay rate of the error probability for increasing $N$.
We repeated this simulation for $4$ different sets of values of $\mathcal{V}_{jx/jy}$, $I_0$, $\chi^2$.

A few technical expedients have been employed to optimize the numerical simulations.
To better sample from the "uncertainty region" of values where the likelihood ratio test is more likely to fail, in order to speed up the numerical simulation, we performed importance sampling from the trinomial distribution generated by the tilted probabilities $Q_i\propto P_{1;i}^sP_{2;i}^{1-s}$, with $s$ numerical solution of the maximization problem of the associated Chernoff exponent.
This technique renders more probable the rare but important events for which the test fails, while adjusting with some weights the bias introduced by the tilted distribution, following the prescription
\begin{equation}
	E_P[f(X)]=E_Q\left[f(X) \frac{P[X]}{Q[X]}\right]=E_Q[f(X) W[X]],
    \label{eq:ExpWeights}
\end{equation}
where $E_D[f(X)]$ is the expectation value of $f(X)$ over the distribution $D(x)$, and $W(x)=P(x)/Q(x)$ are the weights.
In the case at hand, when sampling $x\equiv(M_0,M_1,M_2)$ with hypothesis $j=1,2$, $P(x)\equiv P_j(M_0,M_1,M_2)$ is the trinomial distribution generated by the probabilities $P_{j;i}$,  $Q(x)\equiv Q_j(M_0,M_1,M_2)$ is the trinomial distribution generated by the tilted probabilities $Q_i$, and the weights become
\begin{gather}
\begin{split}
W(x)\equiv W_1(M_0,M_1,M_2)&=\left(\frac{P_{1;0}^{1-s}}{P_{2;0}^{1-s}}\right)^{M_0}\left(\frac{P_{1;1}^{1-s}}{P_{2;1}^{1-s}}\right)^{M_1}\left(\frac{P_{1;2}^{1-s}}{P_{2;2}^{1-s}}\right)^{M_2}\ \mathrm{for}\ j=1,\\
 W(x)\equiv W_2(M_0,M_1,M_2)&=\left(\frac{P_{1;0}^{-s}}{P_{2;0}^{-s}}\right)^{M_0}\left(\frac{P_{1;1}^{-s}}{P_{2;1}^{-s}}\right)^{M_1}\left(\frac{P_{1;2}^{-s}}{P_{2;2}^{-s}}\right)^{M_2}\ \quad\ \mathrm{for}\ j=2.
\end{split}
\end{gather}By employing Eq.~\eqref{eq:ExpWeights}, we can thus evaluate the probability of error $P_e=\frac{1}{2}(E_{P_1}[\boldsymbol{1}_{\Lambda<1}+\boldsymbol{1}_{\Lambda=1}/2]+E_{P_2}[\boldsymbol{1}_{\Lambda>1}+\boldsymbol{1}_{\Lambda=1}/2])$, where $\boldsymbol{1}_{R}$ denotes the characteristic function of the region of values $(M_0,M_1,M_2)\in R$.
After generating $10^4$ values $\{M^{(k)}_0,M^{(k)}_1,M^{(k)}_2\}_{k=1,\dots,10^4}\equiv\{\boldsymbol{M}^{(k)}\}_{k=1,\dots,10^4}$ sampled by the tilted trinomial distribution $Q$, we can retrieve
\begin{equation}
	P_e=\frac{1}{2*10^4}\sum_{k=1}^{10^4} W_1(\boldsymbol{M}^{(k)})\theta[1-\Lambda(\boldsymbol{M}^{(k)})]+W_2(\boldsymbol{M}^{(k)})\theta[\Lambda(\boldsymbol{M}^{(k)})-1],
\end{equation}
where $\theta(x)$ is the Heaviside function such that $\theta(x>0)=1$, $\theta(x<0)=0$ and $\theta(0)=1/2$.
Finally, to tame the exponentially small and large values of the likelihood ratio and of the error probability, we performed all the simulations in the logarithmic domain, evaluating instead
\begin{multline}
	\ln(P_e)=-\ln(2*10^4)+\ln(w^*)\\+\ln\left\{\sum_{k=1}^{10^4}\exp\left[\ln W_1(\boldsymbol{M}^{(k)})-\ln w^*\right]\theta[-\ln\Lambda(\boldsymbol{M}^{(k)})]+\exp\left[\ln W_2(\boldsymbol{M}^{(k)})-\ln w^*\right]\theta[\ln\Lambda(\boldsymbol{M}^{(k)})]\right\},
\end{multline}
where $w^*=\min\limits_{\substack{j=1,2\\k=1,\dots,10^4}}W_j(\boldsymbol{M}^{(k)})$

\bibliography{references.bib}

\begin{thebibliography}{58}%
\makeatletter
\providecommand \@ifxundefined [1]{%
 \@ifx{#1\undefined}
}%
\providecommand \@ifnum [1]{%
 \ifnum #1\expandafter \@firstoftwo
 \else \expandafter \@secondoftwo
 \fi
}%
\providecommand \@ifx [1]{%
 \ifx #1\expandafter \@firstoftwo
 \else \expandafter \@secondoftwo
 \fi
}%
\providecommand \natexlab [1]{#1}%
\providecommand \enquote  [1]{``#1''}%
\providecommand \bibnamefont  [1]{#1}%
\providecommand \bibfnamefont [1]{#1}%
\providecommand \citenamefont [1]{#1}%
\providecommand \href@noop [0]{\@secondoftwo}%
\providecommand \href [0]{\begingroup \@sanitize@url \@href}%
\providecommand \@href[1]{\@@startlink{#1}\@@href}%
\providecommand \@@href[1]{\endgroup#1\@@endlink}%
\providecommand \@sanitize@url [0]{\catcode `\\12\catcode `\$12\catcode `\&12\catcode `\#12\catcode `\^12\catcode `\_12\catcode `\%12\relax}%
\providecommand \@@startlink[1]{}%
\providecommand \@@endlink[0]{}%
\providecommand \url  [0]{\begingroup\@sanitize@url \@url }%
\providecommand \@url [1]{\endgroup\@href {#1}{\urlprefix }}%
\providecommand \urlprefix  [0]{URL }%
\providecommand \Eprint [0]{\href }%
\providecommand \doibase [0]{https://doi.org/}%
\providecommand \selectlanguage [0]{\@gobble}%
\providecommand \bibinfo  [0]{\@secondoftwo}%
\providecommand \bibfield  [0]{\@secondoftwo}%
\providecommand \translation [1]{[#1]}%
\providecommand \BibitemOpen [0]{}%
\providecommand \bibitemStop [0]{}%
\providecommand \bibitemNoStop [0]{.\EOS\space}%
\providecommand \EOS [0]{\spacefactor3000\relax}%
\providecommand \BibitemShut  [1]{\csname bibitem#1\endcsname}%
\let\auto@bib@innerbib\@empty
\bibitem [{\citenamefont {Rayleigh}(1879)}]{Rayleigh1879}%
  \BibitemOpen
  \bibfield  {author} {\bibinfo {author} {\bibnamefont {Rayleigh}},\ }\bibfield  {title} {\bibinfo {title} {Xxxi. investigations in optics, with special reference to the spectroscope},\ }\href {https://doi.org/10.1080/14786447908639684} {\bibfield  {journal} {\bibinfo  {journal} {The London, Edinburgh, and Dublin Philosophical Magazine and Journal of Science}\ }\textbf {\bibinfo {volume} {8}},\ \bibinfo {pages} {261} (\bibinfo {year} {1879})},\ \Eprint {https://arxiv.org/abs/https://doi.org/10.1080/14786447908639684} {https://doi.org/10.1080/14786447908639684} \BibitemShut {NoStop}%
\bibitem [{\citenamefont {Helstrom}(1969)}]{Helstrom1969}%
  \BibitemOpen
  \bibfield  {author} {\bibinfo {author} {\bibfnamefont {C.~W.}\ \bibnamefont {Helstrom}},\ }\bibfield  {title} {\bibinfo {title} {Quantum detection and estimation theory},\ }\href {https://doi.org/10.1007/BF01007479} {\bibfield  {journal} {\bibinfo  {journal} {Journal of Statistical Physics}\ }\textbf {\bibinfo {volume} {1}},\ \bibinfo {pages} {231} (\bibinfo {year} {1969})}\BibitemShut {NoStop}%
\bibitem [{\citenamefont {Nair}\ and\ \citenamefont {Tsang}(2016)}]{Tsang2016L}%
  \BibitemOpen
  \bibfield  {author} {\bibinfo {author} {\bibfnamefont {R.}~\bibnamefont {Nair}}\ and\ \bibinfo {author} {\bibfnamefont {M.}~\bibnamefont {Tsang}},\ }\bibfield  {title} {\bibinfo {title} {Far-field superresolution of thermal electromagnetic sources at the quantum limit},\ }\href {https://doi.org/10.1103/PhysRevLett.117.190801} {\bibfield  {journal} {\bibinfo  {journal} {Phys. Rev. Lett.}\ }\textbf {\bibinfo {volume} {117}},\ \bibinfo {pages} {190801} (\bibinfo {year} {2016})}\BibitemShut {NoStop}%
\bibitem [{\citenamefont {Lupo}\ and\ \citenamefont {Pirandola}(2016)}]{Lupo2016}%
  \BibitemOpen
  \bibfield  {author} {\bibinfo {author} {\bibfnamefont {C.}~\bibnamefont {Lupo}}\ and\ \bibinfo {author} {\bibfnamefont {S.}~\bibnamefont {Pirandola}},\ }\bibfield  {title} {\bibinfo {title} {Ultimate precision bound of quantum and subwavelength imaging},\ }\href {https://doi.org/10.1103/PhysRevLett.117.190802} {\bibfield  {journal} {\bibinfo  {journal} {Phys. Rev. Lett.}\ }\textbf {\bibinfo {volume} {117}},\ \bibinfo {pages} {190802} (\bibinfo {year} {2016})}\BibitemShut {NoStop}%
\bibitem [{\citenamefont {\ifmmode \check{R}\else \v{R}\fi{}eha\ifmmode~\check{c}\else \v{c}\fi{}ek}\ \emph {et~al.}(2017)\citenamefont {\ifmmode \check{R}\else \v{R}\fi{}eha\ifmmode~\check{c}\else \v{c}\fi{}ek}, \citenamefont {Hradil}, \citenamefont {Stoklasa}, \citenamefont {Pa\'ur}, \citenamefont {Grover}, \citenamefont {Krzic},\ and\ \citenamefont {S\'anchez-Soto}}]{Rehacek2017A}%
  \BibitemOpen
  \bibfield  {author} {\bibinfo {author} {\bibfnamefont {J.}~\bibnamefont {\ifmmode \check{R}\else \v{R}\fi{}eha\ifmmode~\check{c}\else \v{c}\fi{}ek}}, \bibinfo {author} {\bibfnamefont {Z.}~\bibnamefont {Hradil}}, \bibinfo {author} {\bibfnamefont {B.}~\bibnamefont {Stoklasa}}, \bibinfo {author} {\bibfnamefont {M.}~\bibnamefont {Pa\'ur}}, \bibinfo {author} {\bibfnamefont {J.}~\bibnamefont {Grover}}, \bibinfo {author} {\bibfnamefont {A.}~\bibnamefont {Krzic}},\ and\ \bibinfo {author} {\bibfnamefont {L.~L.}\ \bibnamefont {S\'anchez-Soto}},\ }\bibfield  {title} {\bibinfo {title} {Multiparameter quantum metrology of incoherent point sources: Towards realistic superresolution},\ }\href {https://doi.org/10.1103/PhysRevA.96.062107} {\bibfield  {journal} {\bibinfo  {journal} {Phys. Rev. A}\ }\textbf {\bibinfo {volume} {96}},\ \bibinfo {pages} {062107} (\bibinfo {year} {2017})}\BibitemShut {NoStop}%
\bibitem [{\citenamefont {Yu}\ and\ \citenamefont {Prasad}(2018)}]{Yu2018}%
  \BibitemOpen
  \bibfield  {author} {\bibinfo {author} {\bibfnamefont {Z.}~\bibnamefont {Yu}}\ and\ \bibinfo {author} {\bibfnamefont {S.}~\bibnamefont {Prasad}},\ }\bibfield  {title} {\bibinfo {title} {Quantum limited superresolution of an incoherent source pair in three dimensions},\ }\href {https://doi.org/10.1103/PhysRevLett.121.180504} {\bibfield  {journal} {\bibinfo  {journal} {Phys. Rev. Lett.}\ }\textbf {\bibinfo {volume} {121}},\ \bibinfo {pages} {180504} (\bibinfo {year} {2018})}\BibitemShut {NoStop}%
\bibitem [{\citenamefont {Tsang}(2019{\natexlab{a}})}]{Tsang2019A}%
  \BibitemOpen
  \bibfield  {author} {\bibinfo {author} {\bibfnamefont {M.}~\bibnamefont {Tsang}},\ }\bibfield  {title} {\bibinfo {title} {Quantum limit to subdiffraction incoherent optical imaging},\ }\href {https://doi.org/10.1103/PhysRevA.99.012305} {\bibfield  {journal} {\bibinfo  {journal} {Phys. Rev. A}\ }\textbf {\bibinfo {volume} {99}},\ \bibinfo {pages} {012305} (\bibinfo {year} {2019}{\natexlab{a}})}\BibitemShut {NoStop}%
\bibitem [{\citenamefont {Zhou}\ and\ \citenamefont {Jiang}(2019)}]{Zhou2019}%
  \BibitemOpen
  \bibfield  {author} {\bibinfo {author} {\bibfnamefont {S.}~\bibnamefont {Zhou}}\ and\ \bibinfo {author} {\bibfnamefont {L.}~\bibnamefont {Jiang}},\ }\bibfield  {title} {\bibinfo {title} {Modern description of rayleigh's criterion},\ }\href {https://doi.org/10.1103/PhysRevA.99.013808} {\bibfield  {journal} {\bibinfo  {journal} {Phys. Rev. A}\ }\textbf {\bibinfo {volume} {99}},\ \bibinfo {pages} {013808} (\bibinfo {year} {2019})}\BibitemShut {NoStop}%
\bibitem [{\citenamefont {Napoli}\ \emph {et~al.}(2019)\citenamefont {Napoli}, \citenamefont {Piano}, \citenamefont {Leach}, \citenamefont {Adesso},\ and\ \citenamefont {Tufarelli}}]{Napoli2019}%
  \BibitemOpen
  \bibfield  {author} {\bibinfo {author} {\bibfnamefont {C.}~\bibnamefont {Napoli}}, \bibinfo {author} {\bibfnamefont {S.}~\bibnamefont {Piano}}, \bibinfo {author} {\bibfnamefont {R.}~\bibnamefont {Leach}}, \bibinfo {author} {\bibfnamefont {G.}~\bibnamefont {Adesso}},\ and\ \bibinfo {author} {\bibfnamefont {T.}~\bibnamefont {Tufarelli}},\ }\bibfield  {title} {\bibinfo {title} {Towards superresolution surface metrology: Quantum estimation of angular and axial separations},\ }\href {https://doi.org/10.1103/PhysRevLett.122.140505} {\bibfield  {journal} {\bibinfo  {journal} {Phys. Rev. Lett.}\ }\textbf {\bibinfo {volume} {122}},\ \bibinfo {pages} {140505} (\bibinfo {year} {2019})}\BibitemShut {NoStop}%
\bibitem [{\citenamefont {Tsang}\ \emph {et~al.}(2016)\citenamefont {Tsang}, \citenamefont {Nair},\ and\ \citenamefont {Lu}}]{Tsang2016X}%
  \BibitemOpen
  \bibfield  {author} {\bibinfo {author} {\bibfnamefont {M.}~\bibnamefont {Tsang}}, \bibinfo {author} {\bibfnamefont {R.}~\bibnamefont {Nair}},\ and\ \bibinfo {author} {\bibfnamefont {X.-M.}\ \bibnamefont {Lu}},\ }\bibfield  {title} {\bibinfo {title} {Quantum theory of superresolution for two incoherent optical point sources},\ }\href {https://doi.org/10.1103/PhysRevX.6.031033} {\bibfield  {journal} {\bibinfo  {journal} {Phys. Rev. X}\ }\textbf {\bibinfo {volume} {6}},\ \bibinfo {pages} {031033} (\bibinfo {year} {2016})}\BibitemShut {NoStop}%
\bibitem [{\citenamefont {Rehacek}\ \emph {et~al.}(2017)\citenamefont {Rehacek}, \citenamefont {Pa\'{u}r}, \citenamefont {Stoklasa}, \citenamefont {Hradil},\ and\ \citenamefont {S\'{a}nchez-Soto}}]{Rehacek2017}%
  \BibitemOpen
  \bibfield  {author} {\bibinfo {author} {\bibfnamefont {J.}~\bibnamefont {Rehacek}}, \bibinfo {author} {\bibfnamefont {M.}~\bibnamefont {Pa\'{u}r}}, \bibinfo {author} {\bibfnamefont {B.}~\bibnamefont {Stoklasa}}, \bibinfo {author} {\bibfnamefont {Z.}~\bibnamefont {Hradil}},\ and\ \bibinfo {author} {\bibfnamefont {L.~L.}\ \bibnamefont {S\'{a}nchez-Soto}},\ }\bibfield  {title} {\bibinfo {title} {Optimal measurements for resolution beyond the rayleigh limit},\ }\href {https://doi.org/10.1364/OL.42.000231} {\bibfield  {journal} {\bibinfo  {journal} {Opt. Lett.}\ }\textbf {\bibinfo {volume} {42}},\ \bibinfo {pages} {231} (\bibinfo {year} {2017})}\BibitemShut {NoStop}%
\bibitem [{\citenamefont {Ang}\ \emph {et~al.}(2017)\citenamefont {Ang}, \citenamefont {Nair},\ and\ \citenamefont {Tsang}}]{Ang2017}%
  \BibitemOpen
  \bibfield  {author} {\bibinfo {author} {\bibfnamefont {S.~Z.}\ \bibnamefont {Ang}}, \bibinfo {author} {\bibfnamefont {R.}~\bibnamefont {Nair}},\ and\ \bibinfo {author} {\bibfnamefont {M.}~\bibnamefont {Tsang}},\ }\bibfield  {title} {\bibinfo {title} {Quantum limit for two-dimensional resolution of two incoherent optical point sources},\ }\href {https://doi.org/10.1103/PhysRevA.95.063847} {\bibfield  {journal} {\bibinfo  {journal} {Phys. Rev. A}\ }\textbf {\bibinfo {volume} {95}},\ \bibinfo {pages} {063847} (\bibinfo {year} {2017})}\BibitemShut {NoStop}%
\bibitem [{\citenamefont {Tsang}(2017)}]{Tsang2017}%
  \BibitemOpen
  \bibfield  {author} {\bibinfo {author} {\bibfnamefont {M.}~\bibnamefont {Tsang}},\ }\bibfield  {title} {\bibinfo {title} {Subdiffraction incoherent optical imaging via spatial-mode demultiplexing},\ }\href {https://doi.org/10.1088/1367-2630/aa60ee} {\bibfield  {journal} {\bibinfo  {journal} {New Journal of Physics}\ }\textbf {\bibinfo {volume} {19}},\ \bibinfo {pages} {023054} (\bibinfo {year} {2017})}\BibitemShut {NoStop}%
\bibitem [{\citenamefont {Tsang}(2018)}]{Tsang2018}%
  \BibitemOpen
  \bibfield  {author} {\bibinfo {author} {\bibfnamefont {M.}~\bibnamefont {Tsang}},\ }\bibfield  {title} {\bibinfo {title} {Subdiffraction incoherent optical imaging via spatial-mode demultiplexing: Semiclassical treatment},\ }\href {https://doi.org/10.1103/PhysRevA.97.023830} {\bibfield  {journal} {\bibinfo  {journal} {Phys. Rev. A}\ }\textbf {\bibinfo {volume} {97}},\ \bibinfo {pages} {023830} (\bibinfo {year} {2018})}\BibitemShut {NoStop}%
\bibitem [{\citenamefont {Dutton}\ \emph {et~al.}(2019)\citenamefont {Dutton}, \citenamefont {Kerviche}, \citenamefont {Ashok},\ and\ \citenamefont {Guha}}]{Dutton2019}%
  \BibitemOpen
  \bibfield  {author} {\bibinfo {author} {\bibfnamefont {Z.}~\bibnamefont {Dutton}}, \bibinfo {author} {\bibfnamefont {R.}~\bibnamefont {Kerviche}}, \bibinfo {author} {\bibfnamefont {A.}~\bibnamefont {Ashok}},\ and\ \bibinfo {author} {\bibfnamefont {S.}~\bibnamefont {Guha}},\ }\bibfield  {title} {\bibinfo {title} {Attaining the quantum limit of superresolution in imaging an object's length via predetection spatial-mode sorting},\ }\href {https://doi.org/10.1103/PhysRevA.99.033847} {\bibfield  {journal} {\bibinfo  {journal} {Phys. Rev. A}\ }\textbf {\bibinfo {volume} {99}},\ \bibinfo {pages} {033847} (\bibinfo {year} {2019})}\BibitemShut {NoStop}%
\bibitem [{\citenamefont {Hervas}\ \emph {et~al.}(2024)\citenamefont {Hervas}, \citenamefont {S\'anchez-Soto}, \citenamefont {Goldberg}, \citenamefont {Hradil},\ and\ \citenamefont {\ifmmode \check{R}\else \v{R}\fi{}eh\'a\ifmmode~\check{c}\else \v{c}\fi{}ek}}]{Hervas2024}%
  \BibitemOpen
  \bibfield  {author} {\bibinfo {author} {\bibfnamefont {J.~R.}\ \bibnamefont {Hervas}}, \bibinfo {author} {\bibfnamefont {L.~L.}\ \bibnamefont {S\'anchez-Soto}}, \bibinfo {author} {\bibfnamefont {A.~Z.}\ \bibnamefont {Goldberg}}, \bibinfo {author} {\bibfnamefont {Z.}~\bibnamefont {Hradil}},\ and\ \bibinfo {author} {\bibfnamefont {J.}~\bibnamefont {\ifmmode \check{R}\else \v{R}\fi{}eh\'a\ifmmode~\check{c}\else \v{c}\fi{}ek}},\ }\bibfield  {title} {\bibinfo {title} {Optimizing measurement tradeoffs in multiparameter spatial superresolution},\ }\href {https://doi.org/10.1103/PhysRevA.110.033716} {\bibfield  {journal} {\bibinfo  {journal} {Phys. Rev. A}\ }\textbf {\bibinfo {volume} {110}},\ \bibinfo {pages} {033716} (\bibinfo {year} {2024})}\BibitemShut {NoStop}%
\bibitem [{\citenamefont {Bisketzi}\ \emph {et~al.}(2019)\citenamefont {Bisketzi}, \citenamefont {Branford},\ and\ \citenamefont {Datta}}]{Bisketzi2019}%
  \BibitemOpen
  \bibfield  {author} {\bibinfo {author} {\bibfnamefont {E.}~\bibnamefont {Bisketzi}}, \bibinfo {author} {\bibfnamefont {D.}~\bibnamefont {Branford}},\ and\ \bibinfo {author} {\bibfnamefont {A.}~\bibnamefont {Datta}},\ }\bibfield  {title} {\bibinfo {title} {Quantum limits of localisation microscopy},\ }\href {https://doi.org/10.1088/1367-2630/ab58a0} {\bibfield  {journal} {\bibinfo  {journal} {New Journal of Physics}\ }\textbf {\bibinfo {volume} {21}},\ \bibinfo {pages} {123032} (\bibinfo {year} {2019})}\BibitemShut {NoStop}%
\bibitem [{\citenamefont {Pushkina}\ \emph {et~al.}(2021)\citenamefont {Pushkina}, \citenamefont {Maltese}, \citenamefont {Costa-Filho}, \citenamefont {Patel},\ and\ \citenamefont {Lvovsky}}]{Pushkina2021}%
  \BibitemOpen
  \bibfield  {author} {\bibinfo {author} {\bibfnamefont {A.~A.}\ \bibnamefont {Pushkina}}, \bibinfo {author} {\bibfnamefont {G.}~\bibnamefont {Maltese}}, \bibinfo {author} {\bibfnamefont {J.~I.}\ \bibnamefont {Costa-Filho}}, \bibinfo {author} {\bibfnamefont {P.}~\bibnamefont {Patel}},\ and\ \bibinfo {author} {\bibfnamefont {A.~I.}\ \bibnamefont {Lvovsky}},\ }\bibfield  {title} {\bibinfo {title} {Superresolution linear optical imaging in the far field},\ }\href {https://doi.org/10.1103/PhysRevLett.127.253602} {\bibfield  {journal} {\bibinfo  {journal} {Phys. Rev. Lett.}\ }\textbf {\bibinfo {volume} {127}},\ \bibinfo {pages} {253602} (\bibinfo {year} {2021})}\BibitemShut {NoStop}%
\bibitem [{\citenamefont {Frank}\ \emph {et~al.}(2023)\citenamefont {Frank}, \citenamefont {Duplinskiy}, \citenamefont {Bearne},\ and\ \citenamefont {Lvovsky}}]{Frank2023}%
  \BibitemOpen
  \bibfield  {author} {\bibinfo {author} {\bibfnamefont {J.}~\bibnamefont {Frank}}, \bibinfo {author} {\bibfnamefont {A.}~\bibnamefont {Duplinskiy}}, \bibinfo {author} {\bibfnamefont {K.}~\bibnamefont {Bearne}},\ and\ \bibinfo {author} {\bibfnamefont {A.~I.}\ \bibnamefont {Lvovsky}},\ }\bibfield  {title} {\bibinfo {title} {Passive superresolution imaging of incoherent objects},\ }\href {https://doi.org/10.1364/OPTICA.493718} {\bibfield  {journal} {\bibinfo  {journal} {Optica}\ }\textbf {\bibinfo {volume} {10}},\ \bibinfo {pages} {1147} (\bibinfo {year} {2023})}\BibitemShut {NoStop}%
\bibitem [{\citenamefont {Santamaria}\ \emph {et~al.}(2023)\citenamefont {Santamaria}, \citenamefont {Pallotti}, \citenamefont {de~Cumis}, \citenamefont {Dequal},\ and\ \citenamefont {Lupo}}]{Santamaria2023}%
  \BibitemOpen
  \bibfield  {author} {\bibinfo {author} {\bibfnamefont {L.}~\bibnamefont {Santamaria}}, \bibinfo {author} {\bibfnamefont {D.}~\bibnamefont {Pallotti}}, \bibinfo {author} {\bibfnamefont {M.~S.}\ \bibnamefont {de~Cumis}}, \bibinfo {author} {\bibfnamefont {D.}~\bibnamefont {Dequal}},\ and\ \bibinfo {author} {\bibfnamefont {C.}~\bibnamefont {Lupo}},\ }\bibfield  {title} {\bibinfo {title} {Spatial-mode demultiplexing for enhanced intensity and distance measurement},\ }\href {https://doi.org/10.1364/OE.486617} {\bibfield  {journal} {\bibinfo  {journal} {Opt. Express}\ }\textbf {\bibinfo {volume} {31}},\ \bibinfo {pages} {33930} (\bibinfo {year} {2023})}\BibitemShut {NoStop}%
\bibitem [{\citenamefont {Wallis}\ \emph {et~al.}(2025)\citenamefont {Wallis}, \citenamefont {Gozzard}, \citenamefont {Frost}, \citenamefont {Collier}, \citenamefont {Maron},\ and\ \citenamefont {Dix-Matthews}}]{Wallis2025}%
  \BibitemOpen
  \bibfield  {author} {\bibinfo {author} {\bibfnamefont {J.~S.}\ \bibnamefont {Wallis}}, \bibinfo {author} {\bibfnamefont {D.~R.}\ \bibnamefont {Gozzard}}, \bibinfo {author} {\bibfnamefont {A.~M.}\ \bibnamefont {Frost}}, \bibinfo {author} {\bibfnamefont {J.~J.}\ \bibnamefont {Collier}}, \bibinfo {author} {\bibfnamefont {N.}~\bibnamefont {Maron}},\ and\ \bibinfo {author} {\bibfnamefont {B.~P.}\ \bibnamefont {Dix-Matthews}},\ }\bibfield  {title} {\bibinfo {title} {Spatial mode demultiplexing for super-resolved source parameter estimation},\ }\href {https://doi.org/10.1364/OE.563503} {\bibfield  {journal} {\bibinfo  {journal} {Opt. Express}\ }\textbf {\bibinfo {volume} {33}},\ \bibinfo {pages} {34651} (\bibinfo {year} {2025})}\BibitemShut {NoStop}%
\bibitem [{\citenamefont {Sorelli}\ \emph {et~al.}(2021)\citenamefont {Sorelli}, \citenamefont {Gessner}, \citenamefont {Walschaers},\ and\ \citenamefont {Treps}}]{Sorelli2021}%
  \BibitemOpen
  \bibfield  {author} {\bibinfo {author} {\bibfnamefont {G.}~\bibnamefont {Sorelli}}, \bibinfo {author} {\bibfnamefont {M.}~\bibnamefont {Gessner}}, \bibinfo {author} {\bibfnamefont {M.}~\bibnamefont {Walschaers}},\ and\ \bibinfo {author} {\bibfnamefont {N.}~\bibnamefont {Treps}},\ }\bibfield  {title} {\bibinfo {title} {Optimal observables and estimators for practical superresolution imaging},\ }\href {https://doi.org/10.1103/PhysRevLett.127.123604} {\bibfield  {journal} {\bibinfo  {journal} {Phys. Rev. Lett.}\ }\textbf {\bibinfo {volume} {127}},\ \bibinfo {pages} {123604} (\bibinfo {year} {2021})}\BibitemShut {NoStop}%
\bibitem [{\citenamefont {Linowski}\ \emph {et~al.}(2023)\citenamefont {Linowski}, \citenamefont {Schlichtholz}, \citenamefont {Sorelli}, \citenamefont {Gessner}, \citenamefont {Walschaers}, \citenamefont {Treps},\ and\ \citenamefont {Rudnicki}}]{Linowski2023}%
  \BibitemOpen
  \bibfield  {author} {\bibinfo {author} {\bibfnamefont {T.}~\bibnamefont {Linowski}}, \bibinfo {author} {\bibfnamefont {K.}~\bibnamefont {Schlichtholz}}, \bibinfo {author} {\bibfnamefont {G.}~\bibnamefont {Sorelli}}, \bibinfo {author} {\bibfnamefont {M.}~\bibnamefont {Gessner}}, \bibinfo {author} {\bibfnamefont {M.}~\bibnamefont {Walschaers}}, \bibinfo {author} {\bibfnamefont {N.}~\bibnamefont {Treps}},\ and\ \bibinfo {author} {\bibfnamefont {L.}~\bibnamefont {Rudnicki}},\ }\bibfield  {title} {\bibinfo {title} {Application range of crosstalk-affected spatial demultiplexing for resolving separations between unbalanced sources},\ }\href {https://doi.org/10.1088/1367-2630/ad0173} {\bibfield  {journal} {\bibinfo  {journal} {New Journal of Physics}\ }\textbf {\bibinfo {volume} {25}},\ \bibinfo {pages} {103050} (\bibinfo {year} {2023})}\BibitemShut {NoStop}%
\bibitem [{\citenamefont {Yang}\ \emph {et~al.}(2016)\citenamefont {Yang}, \citenamefont {Tashchilina}, \citenamefont {Moiseev}, \citenamefont {Simon},\ and\ \citenamefont {Lvovsky}}]{Yang2016}%
  \BibitemOpen
  \bibfield  {author} {\bibinfo {author} {\bibfnamefont {F.}~\bibnamefont {Yang}}, \bibinfo {author} {\bibfnamefont {A.}~\bibnamefont {Tashchilina}}, \bibinfo {author} {\bibfnamefont {E.~S.}\ \bibnamefont {Moiseev}}, \bibinfo {author} {\bibfnamefont {C.}~\bibnamefont {Simon}},\ and\ \bibinfo {author} {\bibfnamefont {A.~I.}\ \bibnamefont {Lvovsky}},\ }\bibfield  {title} {\bibinfo {title} {Far-field linear optical superresolution via heterodyne detection in a higher-order local oscillator mode},\ }\href {https://doi.org/10.1364/OPTICA.3.001148} {\bibfield  {journal} {\bibinfo  {journal} {Optica}\ }\textbf {\bibinfo {volume} {3}},\ \bibinfo {pages} {1148} (\bibinfo {year} {2016})}\BibitemShut {NoStop}%
\bibitem [{\citenamefont {Yang}\ \emph {et~al.}(2017)\citenamefont {Yang}, \citenamefont {Nair}, \citenamefont {Tsang}, \citenamefont {Simon},\ and\ \citenamefont {Lvovsky}}]{Yang2017}%
  \BibitemOpen
  \bibfield  {author} {\bibinfo {author} {\bibfnamefont {F.}~\bibnamefont {Yang}}, \bibinfo {author} {\bibfnamefont {R.}~\bibnamefont {Nair}}, \bibinfo {author} {\bibfnamefont {M.}~\bibnamefont {Tsang}}, \bibinfo {author} {\bibfnamefont {C.}~\bibnamefont {Simon}},\ and\ \bibinfo {author} {\bibfnamefont {A.~I.}\ \bibnamefont {Lvovsky}},\ }\bibfield  {title} {\bibinfo {title} {Fisher information for far-field linear optical superresolution via homodyne or heterodyne detection in a higher-order local oscillator mode},\ }\href {https://doi.org/10.1103/PhysRevA.96.063829} {\bibfield  {journal} {\bibinfo  {journal} {Phys. Rev. A}\ }\textbf {\bibinfo {volume} {96}},\ \bibinfo {pages} {063829} (\bibinfo {year} {2017})}\BibitemShut {NoStop}%
\bibitem [{\citenamefont {Tham}\ \emph {et~al.}(2017)\citenamefont {Tham}, \citenamefont {Ferretti},\ and\ \citenamefont {Steinberg}}]{Tham2017}%
  \BibitemOpen
  \bibfield  {author} {\bibinfo {author} {\bibfnamefont {W.-K.}\ \bibnamefont {Tham}}, \bibinfo {author} {\bibfnamefont {H.}~\bibnamefont {Ferretti}},\ and\ \bibinfo {author} {\bibfnamefont {A.~M.}\ \bibnamefont {Steinberg}},\ }\bibfield  {title} {\bibinfo {title} {Beating rayleigh's curse by imaging using phase information},\ }\href {https://doi.org/10.1103/PhysRevLett.118.070801} {\bibfield  {journal} {\bibinfo  {journal} {Phys. Rev. Lett.}\ }\textbf {\bibinfo {volume} {118}},\ \bibinfo {pages} {070801} (\bibinfo {year} {2017})}\BibitemShut {NoStop}%
\bibitem [{\citenamefont {Bonsma-Fisher}\ \emph {et~al.}(2019)\citenamefont {Bonsma-Fisher}, \citenamefont {Tham}, \citenamefont {Ferretti},\ and\ \citenamefont {Steinberg}}]{Bonsma-Fisher2019}%
  \BibitemOpen
  \bibfield  {author} {\bibinfo {author} {\bibfnamefont {K.~A.~G.}\ \bibnamefont {Bonsma-Fisher}}, \bibinfo {author} {\bibfnamefont {W.-K.}\ \bibnamefont {Tham}}, \bibinfo {author} {\bibfnamefont {H.}~\bibnamefont {Ferretti}},\ and\ \bibinfo {author} {\bibfnamefont {A.~M.}\ \bibnamefont {Steinberg}},\ }\bibfield  {title} {\bibinfo {title} {Realistic sub-rayleigh imaging with phase-sensitive measurements},\ }\href {https://doi.org/10.1088/1367-2630/ab3d97} {\bibfield  {journal} {\bibinfo  {journal} {New Journal of Physics}\ }\textbf {\bibinfo {volume} {21}},\ \bibinfo {pages} {093010} (\bibinfo {year} {2019})}\BibitemShut {NoStop}%
\bibitem [{\citenamefont {Darji}\ \emph {et~al.}(2024)\citenamefont {Darji}, \citenamefont {Kumar},\ and\ \citenamefont {Huang}}]{Darji2024}%
  \BibitemOpen
  \bibfield  {author} {\bibinfo {author} {\bibfnamefont {I.}~\bibnamefont {Darji}}, \bibinfo {author} {\bibfnamefont {S.}~\bibnamefont {Kumar}},\ and\ \bibinfo {author} {\bibfnamefont {Y.-P.}\ \bibnamefont {Huang}},\ }\bibfield  {title} {\bibinfo {title} {Robust super-resolution classifier by nonlinear optics},\ }\href {https://doi.org/10.1364/OL.537295} {\bibfield  {journal} {\bibinfo  {journal} {Opt. Lett.}\ }\textbf {\bibinfo {volume} {49}},\ \bibinfo {pages} {5419} (\bibinfo {year} {2024})}\BibitemShut {NoStop}%
\bibitem [{\citenamefont {Parniak}\ \emph {et~al.}(2018)\citenamefont {Parniak}, \citenamefont {Bor\'owka}, \citenamefont {Boroszko}, \citenamefont {Wasilewski}, \citenamefont {Banaszek},\ and\ \citenamefont {Demkowicz-Dobrza\ifmmode~\acute{n}\else \'{n}\fi{}ski}}]{Parniak2018}%
  \BibitemOpen
  \bibfield  {author} {\bibinfo {author} {\bibfnamefont {M.}~\bibnamefont {Parniak}}, \bibinfo {author} {\bibfnamefont {S.}~\bibnamefont {Bor\'owka}}, \bibinfo {author} {\bibfnamefont {K.}~\bibnamefont {Boroszko}}, \bibinfo {author} {\bibfnamefont {W.}~\bibnamefont {Wasilewski}}, \bibinfo {author} {\bibfnamefont {K.}~\bibnamefont {Banaszek}},\ and\ \bibinfo {author} {\bibfnamefont {R.}~\bibnamefont {Demkowicz-Dobrza\ifmmode~\acute{n}\else \'{n}\fi{}ski}},\ }\bibfield  {title} {\bibinfo {title} {Beating the rayleigh limit using two-photon interference},\ }\href {https://doi.org/10.1103/PhysRevLett.121.250503} {\bibfield  {journal} {\bibinfo  {journal} {Phys. Rev. Lett.}\ }\textbf {\bibinfo {volume} {121}},\ \bibinfo {pages} {250503} (\bibinfo {year} {2018})}\BibitemShut {NoStop}%
\bibitem [{\citenamefont {Muratore}\ \emph {et~al.}(2025)\citenamefont {Muratore}, \citenamefont {Triggiani},\ and\ \citenamefont {Tamma}}]{Muratore2025}%
  \BibitemOpen
  \bibfield  {author} {\bibinfo {author} {\bibfnamefont {S.}~\bibnamefont {Muratore}}, \bibinfo {author} {\bibfnamefont {D.}~\bibnamefont {Triggiani}},\ and\ \bibinfo {author} {\bibfnamefont {V.}~\bibnamefont {Tamma}},\ }\bibfield  {title} {\bibinfo {title} {Superresolution imaging of two incoherent sources via two-photon-interference sampling measurements of the transverse momenta},\ }\href {https://doi.org/10.1103/PhysRevApplied.23.054033} {\bibfield  {journal} {\bibinfo  {journal} {Phys. Rev. Appl.}\ }\textbf {\bibinfo {volume} {23}},\ \bibinfo {pages} {054033} (\bibinfo {year} {2025})}\BibitemShut {NoStop}%
\bibitem [{\citenamefont {Tsang}(2019{\natexlab{b}})}]{Tsang2019}%
  \BibitemOpen
  \bibfield  {author} {\bibinfo {author} {\bibfnamefont {M.}~\bibnamefont {Tsang}},\ }\bibfield  {title} {\bibinfo {title} {Resolving starlight: a quantum perspective},\ }\href {https://doi.org/10.1080/00107514.2020.1736375} {\bibfield  {journal} {\bibinfo  {journal} {Contemporary Physics}\ }\textbf {\bibinfo {volume} {60}},\ \bibinfo {pages} {279} (\bibinfo {year} {2019}{\natexlab{b}})},\ \Eprint {https://arxiv.org/abs/https://doi.org/10.1080/00107514.2020.1736375} {https://doi.org/10.1080/00107514.2020.1736375} \BibitemShut {NoStop}%
\bibitem [{\citenamefont {Defienne}\ \emph {et~al.}(2024)\citenamefont {Defienne}, \citenamefont {Bowen}, \citenamefont {Chekhova}, \citenamefont {Lemos}, \citenamefont {Oron}, \citenamefont {Ramelow}, \citenamefont {Treps},\ and\ \citenamefont {Faccio}}]{Defienne2024}%
  \BibitemOpen
  \bibfield  {author} {\bibinfo {author} {\bibfnamefont {H.}~\bibnamefont {Defienne}}, \bibinfo {author} {\bibfnamefont {W.~P.}\ \bibnamefont {Bowen}}, \bibinfo {author} {\bibfnamefont {M.}~\bibnamefont {Chekhova}}, \bibinfo {author} {\bibfnamefont {G.~B.}\ \bibnamefont {Lemos}}, \bibinfo {author} {\bibfnamefont {D.}~\bibnamefont {Oron}}, \bibinfo {author} {\bibfnamefont {S.}~\bibnamefont {Ramelow}}, \bibinfo {author} {\bibfnamefont {N.}~\bibnamefont {Treps}},\ and\ \bibinfo {author} {\bibfnamefont {D.}~\bibnamefont {Faccio}},\ }\bibfield  {title} {\bibinfo {title} {Advances in quantum imaging},\ }\href {https://doi.org/10.1038/s41566-024-01516-w} {\bibfield  {journal} {\bibinfo  {journal} {Nature Photonics}\ }\textbf {\bibinfo {volume} {18}},\ \bibinfo {pages} {1024} (\bibinfo {year} {2024})}\BibitemShut {NoStop}%
\bibitem [{\citenamefont {Helstrom}(1973)}]{Helstrom1973}%
  \BibitemOpen
  \bibfield  {author} {\bibinfo {author} {\bibfnamefont {C.}~\bibnamefont {Helstrom}},\ }\bibfield  {title} {\bibinfo {title} {Resolution of point sources of light as analyzed by quantum detection theory},\ }\href {https://doi.org/10.1109/TIT.1973.1055052} {\bibfield  {journal} {\bibinfo  {journal} {IEEE Transactions on Information Theory}\ }\textbf {\bibinfo {volume} {19}},\ \bibinfo {pages} {389} (\bibinfo {year} {1973})}\BibitemShut {NoStop}%
\bibitem [{\citenamefont {Audenaert}\ \emph {et~al.}(2007)\citenamefont {Audenaert}, \citenamefont {Calsamiglia}, \citenamefont {Mu\~noz Tapia}, \citenamefont {Bagan}, \citenamefont {Masanes}, \citenamefont {Acin},\ and\ \citenamefont {Verstraete}}]{Audenaert2007}%
  \BibitemOpen
  \bibfield  {author} {\bibinfo {author} {\bibfnamefont {K.~M.~R.}\ \bibnamefont {Audenaert}}, \bibinfo {author} {\bibfnamefont {J.}~\bibnamefont {Calsamiglia}}, \bibinfo {author} {\bibfnamefont {R.}~\bibnamefont {Mu\~noz Tapia}}, \bibinfo {author} {\bibfnamefont {E.}~\bibnamefont {Bagan}}, \bibinfo {author} {\bibfnamefont {L.}~\bibnamefont {Masanes}}, \bibinfo {author} {\bibfnamefont {A.}~\bibnamefont {Acin}},\ and\ \bibinfo {author} {\bibfnamefont {F.}~\bibnamefont {Verstraete}},\ }\bibfield  {title} {\bibinfo {title} {Discriminating states: The quantum chernoff bound},\ }\href {https://doi.org/10.1103/PhysRevLett.98.160501} {\bibfield  {journal} {\bibinfo  {journal} {Phys. Rev. Lett.}\ }\textbf {\bibinfo {volume} {98}},\ \bibinfo {pages} {160501} (\bibinfo {year} {2007})}\BibitemShut {NoStop}%
\bibitem [{\citenamefont {Lu}\ \emph {et~al.}(2018)\citenamefont {Lu}, \citenamefont {Krovi}, \citenamefont {Nair}, \citenamefont {Guha},\ and\ \citenamefont {Shapiro}}]{Lu2018}%
  \BibitemOpen
  \bibfield  {author} {\bibinfo {author} {\bibfnamefont {X.-M.}\ \bibnamefont {Lu}}, \bibinfo {author} {\bibfnamefont {H.}~\bibnamefont {Krovi}}, \bibinfo {author} {\bibfnamefont {R.}~\bibnamefont {Nair}}, \bibinfo {author} {\bibfnamefont {S.}~\bibnamefont {Guha}},\ and\ \bibinfo {author} {\bibfnamefont {J.~H.}\ \bibnamefont {Shapiro}},\ }\bibfield  {title} {\bibinfo {title} {Quantum-optimal detection of one-versus-two incoherent optical sources with arbitrary separation},\ }\href {https://doi.org/10.1038/s41534-018-0114-y} {\bibfield  {journal} {\bibinfo  {journal} {npj Quantum Information}\ }\textbf {\bibinfo {volume} {4}},\ \bibinfo {pages} {64} (\bibinfo {year} {2018})}\BibitemShut {NoStop}%
\bibitem [{\citenamefont {Huang}\ and\ \citenamefont {Lupo}(2021)}]{Huang2021}%
  \BibitemOpen
  \bibfield  {author} {\bibinfo {author} {\bibfnamefont {Z.}~\bibnamefont {Huang}}\ and\ \bibinfo {author} {\bibfnamefont {C.}~\bibnamefont {Lupo}},\ }\bibfield  {title} {\bibinfo {title} {Quantum hypothesis testing for exoplanet detection},\ }\href {https://doi.org/10.1103/PhysRevLett.127.130502} {\bibfield  {journal} {\bibinfo  {journal} {Phys. Rev. Lett.}\ }\textbf {\bibinfo {volume} {127}},\ \bibinfo {pages} {130502} (\bibinfo {year} {2021})}\BibitemShut {NoStop}%
\bibitem [{\citenamefont {Jha}\ \emph {et~al.}(2025)\citenamefont {Jha}, \citenamefont {Fleming}, \citenamefont {Deshler}, \citenamefont {Sajjad}, \citenamefont {Neifeld}, \citenamefont {Ashok},\ and\ \citenamefont {Guha}}]{Jha2025}%
  \BibitemOpen
  \bibfield  {author} {\bibinfo {author} {\bibfnamefont {A.~K.}\ \bibnamefont {Jha}}, \bibinfo {author} {\bibfnamefont {S.}~\bibnamefont {Fleming}}, \bibinfo {author} {\bibfnamefont {N.}~\bibnamefont {Deshler}}, \bibinfo {author} {\bibfnamefont {A.}~\bibnamefont {Sajjad}}, \bibinfo {author} {\bibfnamefont {M.}~\bibnamefont {Neifeld}}, \bibinfo {author} {\bibfnamefont {A.}~\bibnamefont {Ashok}},\ and\ \bibinfo {author} {\bibfnamefont {S.}~\bibnamefont {Guha}},\ }\bibfield  {title} {\bibinfo {title} {Multi-aperture telescopes at the quantum limit of superresolution imaging : Detecting subrayleigh object near a star},\ }\href {https://doi.org/https://doi.org/10.1016/j.actaastro.2024.09.064} {\bibfield  {journal} {\bibinfo  {journal} {Acta Astronautica}\ }\textbf {\bibinfo {volume} {226}},\ \bibinfo {pages} {531} (\bibinfo {year} {2025})}\BibitemShut {NoStop}%
\bibitem [{\citenamefont {Zanforlin}\ \emph {et~al.}(2022)\citenamefont {Zanforlin}, \citenamefont {Lupo}, \citenamefont {Connolly}, \citenamefont {Kok}, \citenamefont {Buller},\ and\ \citenamefont {Huang}}]{Zanforlin2022}%
  \BibitemOpen
  \bibfield  {author} {\bibinfo {author} {\bibfnamefont {U.}~\bibnamefont {Zanforlin}}, \bibinfo {author} {\bibfnamefont {C.}~\bibnamefont {Lupo}}, \bibinfo {author} {\bibfnamefont {P.~W.~R.}\ \bibnamefont {Connolly}}, \bibinfo {author} {\bibfnamefont {P.}~\bibnamefont {Kok}}, \bibinfo {author} {\bibfnamefont {G.~S.}\ \bibnamefont {Buller}},\ and\ \bibinfo {author} {\bibfnamefont {Z.}~\bibnamefont {Huang}},\ }\bibfield  {title} {\bibinfo {title} {Optical quantum super-resolution imaging and hypothesis testing},\ }\href {https://doi.org/10.1038/s41467-022-32977-8} {\bibfield  {journal} {\bibinfo  {journal} {Nature Communications}\ }\textbf {\bibinfo {volume} {13}},\ \bibinfo {pages} {5373} (\bibinfo {year} {2022})}\BibitemShut {NoStop}%
\bibitem [{\citenamefont {Santamaria}\ \emph {et~al.}(2024)\citenamefont {Santamaria}, \citenamefont {Sgobba},\ and\ \citenamefont {Lupo}}]{Santamaria2024}%
  \BibitemOpen
  \bibfield  {author} {\bibinfo {author} {\bibfnamefont {L.}~\bibnamefont {Santamaria}}, \bibinfo {author} {\bibfnamefont {F.}~\bibnamefont {Sgobba}},\ and\ \bibinfo {author} {\bibfnamefont {C.}~\bibnamefont {Lupo}},\ }\bibfield  {title} {\bibinfo {title} {Single-photon sub-rayleigh precision measurements of a pair of incoherent sources of unequal intensity},\ }\href {https://doi.org/10.1364/OPTICAQ.505457} {\bibfield  {journal} {\bibinfo  {journal} {Optica Quantum}\ }\textbf {\bibinfo {volume} {2}},\ \bibinfo {pages} {46} (\bibinfo {year} {2024})}\BibitemShut {NoStop}%
\bibitem [{\citenamefont {Gessner}\ \emph {et~al.}(2020)\citenamefont {Gessner}, \citenamefont {Fabre},\ and\ \citenamefont {Treps}}]{Gessner2020}%
  \BibitemOpen
  \bibfield  {author} {\bibinfo {author} {\bibfnamefont {M.}~\bibnamefont {Gessner}}, \bibinfo {author} {\bibfnamefont {C.}~\bibnamefont {Fabre}},\ and\ \bibinfo {author} {\bibfnamefont {N.}~\bibnamefont {Treps}},\ }\bibfield  {title} {\bibinfo {title} {Superresolution limits from measurement crosstalk},\ }\href {https://doi.org/10.1103/PhysRevLett.125.100501} {\bibfield  {journal} {\bibinfo  {journal} {Phys. Rev. Lett.}\ }\textbf {\bibinfo {volume} {125}},\ \bibinfo {pages} {100501} (\bibinfo {year} {2020})}\BibitemShut {NoStop}%
\bibitem [{\citenamefont {de~Almeida}\ \emph {et~al.}(2021)\citenamefont {de~Almeida}, \citenamefont {Ko\l{}ody\ifmmode~\acute{n}\else \'{n}\fi{}ski}, \citenamefont {Hirche}, \citenamefont {Lewenstein},\ and\ \citenamefont {Skotiniotis}}]{DeAlmeida2021}%
  \BibitemOpen
  \bibfield  {author} {\bibinfo {author} {\bibfnamefont {J.~O.}\ \bibnamefont {de~Almeida}}, \bibinfo {author} {\bibfnamefont {J.}~\bibnamefont {Ko\l{}ody\ifmmode~\acute{n}\else \'{n}\fi{}ski}}, \bibinfo {author} {\bibfnamefont {C.}~\bibnamefont {Hirche}}, \bibinfo {author} {\bibfnamefont {M.}~\bibnamefont {Lewenstein}},\ and\ \bibinfo {author} {\bibfnamefont {M.}~\bibnamefont {Skotiniotis}},\ }\bibfield  {title} {\bibinfo {title} {Discrimination and estimation of incoherent sources under misalignment},\ }\href {https://doi.org/10.1103/PhysRevA.103.022406} {\bibfield  {journal} {\bibinfo  {journal} {Phys. Rev. A}\ }\textbf {\bibinfo {volume} {103}},\ \bibinfo {pages} {022406} (\bibinfo {year} {2021})}\BibitemShut {NoStop}%
\bibitem [{\citenamefont {Schlichtholz}\ \emph {et~al.}(2024)\citenamefont {Schlichtholz}, \citenamefont {Linowski}, \citenamefont {Walschaers}, \citenamefont {Treps}, \citenamefont {Rudnicki},\ and\ \citenamefont {Sorelli}}]{Schlichtholz2024}%
  \BibitemOpen
  \bibfield  {author} {\bibinfo {author} {\bibfnamefont {K.}~\bibnamefont {Schlichtholz}}, \bibinfo {author} {\bibfnamefont {T.}~\bibnamefont {Linowski}}, \bibinfo {author} {\bibfnamefont {M.}~\bibnamefont {Walschaers}}, \bibinfo {author} {\bibfnamefont {N.}~\bibnamefont {Treps}}, \bibinfo {author} {\bibfnamefont {{\L}.}~\bibnamefont {Rudnicki}},\ and\ \bibinfo {author} {\bibfnamefont {G.}~\bibnamefont {Sorelli}},\ }\bibfield  {title} {\bibinfo {title} {Practical tests for sub-rayleigh source discriminations with imperfect demultiplexers},\ }\href {https://doi.org/10.1364/OPTICAQ.502459} {\bibfield  {journal} {\bibinfo  {journal} {Optica Quantum}\ }\textbf {\bibinfo {volume} {2}},\ \bibinfo {pages} {29} (\bibinfo {year} {2024})}\BibitemShut {NoStop}%
\bibitem [{\citenamefont {Linowski}\ \emph {et~al.}(2025)\citenamefont {Linowski}, \citenamefont {Schlichtholz},\ and\ \citenamefont {Sorelli}}]{Linowski2025}%
  \BibitemOpen
  \bibfield  {author} {\bibinfo {author} {\bibfnamefont {T.}~\bibnamefont {Linowski}}, \bibinfo {author} {\bibfnamefont {K.}~\bibnamefont {Schlichtholz}},\ and\ \bibinfo {author} {\bibfnamefont {G.}~\bibnamefont {Sorelli}},\ }\href {https://arxiv.org/abs/2505.00064} {\bibinfo {title} {Quantum-inspired exoplanet detection in the presence of experimental imperfections}} (\bibinfo {year} {2025}),\ \Eprint {https://arxiv.org/abs/2505.00064} {arXiv:2505.00064 [astro-ph.IM]} \BibitemShut {NoStop}%
\bibitem [{\citenamefont {Zhang}\ \emph {et~al.}(2025)\citenamefont {Zhang}, \citenamefont {Zhang}, \citenamefont {Jia}, \citenamefont {Li},\ and\ \citenamefont {Wang}}]{Zhang2025}%
  \BibitemOpen
  \bibfield  {author} {\bibinfo {author} {\bibfnamefont {J.-D.}\ \bibnamefont {Zhang}}, \bibinfo {author} {\bibfnamefont {M.-M.}\ \bibnamefont {Zhang}}, \bibinfo {author} {\bibfnamefont {F.}~\bibnamefont {Jia}}, \bibinfo {author} {\bibfnamefont {C.}~\bibnamefont {Li}},\ and\ \bibinfo {author} {\bibfnamefont {S.}~\bibnamefont {Wang}},\ }\bibfield  {title} {\bibinfo {title} {Quantum-optimal hypothesis testing for discriminating partially coherent optical sources},\ }\href {https://doi.org/10.1103/PhysRevA.111.023706} {\bibfield  {journal} {\bibinfo  {journal} {Phys. Rev. A}\ }\textbf {\bibinfo {volume} {111}},\ \bibinfo {pages} {023706} (\bibinfo {year} {2025})}\BibitemShut {NoStop}%
\bibitem [{\citenamefont {Zhang}\ \emph {et~al.}(2024)\citenamefont {Zhang}, \citenamefont {Zhang}, \citenamefont {Li},\ and\ \citenamefont {Wang}}]{Zhang2024}%
  \BibitemOpen
  \bibfield  {author} {\bibinfo {author} {\bibfnamefont {J.-D.}\ \bibnamefont {Zhang}}, \bibinfo {author} {\bibfnamefont {M.-M.}\ \bibnamefont {Zhang}}, \bibinfo {author} {\bibfnamefont {C.}~\bibnamefont {Li}},\ and\ \bibinfo {author} {\bibfnamefont {S.}~\bibnamefont {Wang}},\ }\bibfield  {title} {\bibinfo {title} {Performance advantage of discriminating one-versus-two incoherent sources based on quantum hypothesis testing},\ }\href {https://doi.org/10.1103/PhysRevA.110.052602} {\bibfield  {journal} {\bibinfo  {journal} {Phys. Rev. A}\ }\textbf {\bibinfo {volume} {110}},\ \bibinfo {pages} {052602} (\bibinfo {year} {2024})}\BibitemShut {NoStop}%
\bibitem [{\citenamefont {Grace}\ and\ \citenamefont {Guha}(2022)}]{Grace2022}%
  \BibitemOpen
  \bibfield  {author} {\bibinfo {author} {\bibfnamefont {M.~R.}\ \bibnamefont {Grace}}\ and\ \bibinfo {author} {\bibfnamefont {S.}~\bibnamefont {Guha}},\ }\bibfield  {title} {\bibinfo {title} {Identifying objects at the quantum limit for superresolution imaging},\ }\href {https://doi.org/10.1103/PhysRevLett.129.180502} {\bibfield  {journal} {\bibinfo  {journal} {Phys. Rev. Lett.}\ }\textbf {\bibinfo {volume} {129}},\ \bibinfo {pages} {180502} (\bibinfo {year} {2022})}\BibitemShut {NoStop}%
\bibitem [{\citenamefont {Buonaiuto}\ and\ \citenamefont {Lupo}(2025)}]{Buonaiuto2025}%
  \BibitemOpen
  \bibfield  {author} {\bibinfo {author} {\bibfnamefont {G.}~\bibnamefont {Buonaiuto}}\ and\ \bibinfo {author} {\bibfnamefont {C.}~\bibnamefont {Lupo}},\ }\bibfield  {title} {\bibinfo {title} {Machine learning with sub-diffraction resolution in the photon-counting regime},\ }\href {https://doi.org/10.1007/s42484-025-00262-8} {\bibfield  {journal} {\bibinfo  {journal} {Quantum Machine Intelligence}\ }\textbf {\bibinfo {volume} {7}},\ \bibinfo {pages} {28} (\bibinfo {year} {2025})}\BibitemShut {NoStop}%
\bibitem [{\citenamefont {Weedbrook}\ \emph {et~al.}(2012)\citenamefont {Weedbrook}, \citenamefont {Pirandola}, \citenamefont {Garc\'{\i}a-Patr\'on}, \citenamefont {Cerf}, \citenamefont {Ralph}, \citenamefont {Shapiro},\ and\ \citenamefont {Lloyd}}]{Weedbrook2012}%
  \BibitemOpen
  \bibfield  {author} {\bibinfo {author} {\bibfnamefont {C.}~\bibnamefont {Weedbrook}}, \bibinfo {author} {\bibfnamefont {S.}~\bibnamefont {Pirandola}}, \bibinfo {author} {\bibfnamefont {R.}~\bibnamefont {Garc\'{\i}a-Patr\'on}}, \bibinfo {author} {\bibfnamefont {N.~J.}\ \bibnamefont {Cerf}}, \bibinfo {author} {\bibfnamefont {T.~C.}\ \bibnamefont {Ralph}}, \bibinfo {author} {\bibfnamefont {J.~H.}\ \bibnamefont {Shapiro}},\ and\ \bibinfo {author} {\bibfnamefont {S.}~\bibnamefont {Lloyd}},\ }\bibfield  {title} {\bibinfo {title} {Gaussian quantum information},\ }\href {https://doi.org/10.1103/RevModPhys.84.621} {\bibfield  {journal} {\bibinfo  {journal} {Rev. Mod. Phys.}\ }\textbf {\bibinfo {volume} {84}},\ \bibinfo {pages} {621} (\bibinfo {year} {2012})}\BibitemShut {NoStop}%
\bibitem [{\citenamefont {Conlon}\ \emph {et~al.}(2023)\citenamefont {Conlon}, \citenamefont {Eilenberger}, \citenamefont {Lam},\ and\ \citenamefont {Assad}}]{Conlon2023}%
  \BibitemOpen
  \bibfield  {author} {\bibinfo {author} {\bibfnamefont {L.~O.}\ \bibnamefont {Conlon}}, \bibinfo {author} {\bibfnamefont {F.}~\bibnamefont {Eilenberger}}, \bibinfo {author} {\bibfnamefont {P.~K.}\ \bibnamefont {Lam}},\ and\ \bibinfo {author} {\bibfnamefont {S.~M.}\ \bibnamefont {Assad}},\ }\bibfield  {title} {\bibinfo {title} {Discriminating mixed qubit states with collective measurements},\ }\href {https://doi.org/10.1038/s42005-023-01454-z} {\bibfield  {journal} {\bibinfo  {journal} {Communications Physics}\ }\textbf {\bibinfo {volume} {6}},\ \bibinfo {pages} {337} (\bibinfo {year} {2023})}\BibitemShut {NoStop}%
\bibitem [{\citenamefont {Farahani}\ and\ \citenamefont {Monteith}(2016)}]{Farahani2016}%
  \BibitemOpen
  \bibfield  {author} {\bibinfo {author} {\bibfnamefont {N.}~\bibnamefont {Farahani}}\ and\ \bibinfo {author} {\bibfnamefont {C.~E.}\ \bibnamefont {Monteith}},\ }\bibfield  {title} {\bibinfo {title} {The coming paradigm shift: A transition from manual to automated microscopy},\ }\href {https://doi.org/https://doi.org/10.4103/2153-3539.189698} {\bibfield  {journal} {\bibinfo  {journal} {Journal of Pathology Informatics}\ }\textbf {\bibinfo {volume} {7}},\ \bibinfo {pages} {35} (\bibinfo {year} {2016})}\BibitemShut {NoStop}%
\bibitem [{\citenamefont {Robertson}\ \emph {et~al.}(2023)\citenamefont {Robertson}, \citenamefont {Tacchella}, \citenamefont {Johnson}, \citenamefont {Hainline}, \citenamefont {Whitler}, \citenamefont {Eisenstein}, \citenamefont {Endsley}, \citenamefont {Rieke}, \citenamefont {Stark}, \citenamefont {Alberts}, \citenamefont {Dressler}, \citenamefont {Egami}, \citenamefont {Hausen}, \citenamefont {Rieke}, \citenamefont {Shivaei}, \citenamefont {Williams}, \citenamefont {Willmer}, \citenamefont {Arribas}, \citenamefont {Bonaventura}, \citenamefont {Bunker}, \citenamefont {Cameron}, \citenamefont {Carniani}, \citenamefont {Charlot}, \citenamefont {Chevallard}, \citenamefont {Curti}, \citenamefont {Curtis-Lake}, \citenamefont {D'Eugenio}, \citenamefont {Jakobsen}, \citenamefont {Looser}, \citenamefont {L{\"u}tzgendorf}, \citenamefont {Maiolino}, \citenamefont {Maseda}, \citenamefont {Rawle}, \citenamefont {Rix}, \citenamefont {Smit}, \citenamefont {{\"U}bler}, \citenamefont {Willott}, \citenamefont {Witstok},
  \citenamefont {Baum}, \citenamefont {Bhatawdekar}, \citenamefont {Boyett}, \citenamefont {Chen}, \citenamefont {de~Graaff}, \citenamefont {Florian}, \citenamefont {Helton}, \citenamefont {Hviding}, \citenamefont {Ji}, \citenamefont {Kumari}, \citenamefont {Lyu}, \citenamefont {Nelson}, \citenamefont {Sandles}, \citenamefont {Saxena}, \citenamefont {Suess}, \citenamefont {Sun}, \citenamefont {Topping},\ and\ \citenamefont {Wallace}}]{Robertson2023}%
  \BibitemOpen
  \bibfield  {author} {\bibinfo {author} {\bibfnamefont {B.~E.}\ \bibnamefont {Robertson}}, \bibinfo {author} {\bibfnamefont {S.}~\bibnamefont {Tacchella}}, \bibinfo {author} {\bibfnamefont {B.~D.}\ \bibnamefont {Johnson}}, \bibinfo {author} {\bibfnamefont {K.}~\bibnamefont {Hainline}}, \bibinfo {author} {\bibfnamefont {L.}~\bibnamefont {Whitler}}, \bibinfo {author} {\bibfnamefont {D.~J.}\ \bibnamefont {Eisenstein}}, \bibinfo {author} {\bibfnamefont {R.}~\bibnamefont {Endsley}}, \bibinfo {author} {\bibfnamefont {M.}~\bibnamefont {Rieke}}, \bibinfo {author} {\bibfnamefont {D.~P.}\ \bibnamefont {Stark}}, \bibinfo {author} {\bibfnamefont {S.}~\bibnamefont {Alberts}}, \bibinfo {author} {\bibfnamefont {A.}~\bibnamefont {Dressler}}, \bibinfo {author} {\bibfnamefont {E.}~\bibnamefont {Egami}}, \bibinfo {author} {\bibfnamefont {R.}~\bibnamefont {Hausen}}, \bibinfo {author} {\bibfnamefont {G.}~\bibnamefont {Rieke}}, \bibinfo {author} {\bibfnamefont {I.}~\bibnamefont {Shivaei}}, \bibinfo {author} {\bibfnamefont
  {C.~C.}\ \bibnamefont {Williams}}, \bibinfo {author} {\bibfnamefont {C.~N.~A.}\ \bibnamefont {Willmer}}, \bibinfo {author} {\bibfnamefont {S.}~\bibnamefont {Arribas}}, \bibinfo {author} {\bibfnamefont {N.}~\bibnamefont {Bonaventura}}, \bibinfo {author} {\bibfnamefont {A.}~\bibnamefont {Bunker}}, \bibinfo {author} {\bibfnamefont {A.~J.}\ \bibnamefont {Cameron}}, \bibinfo {author} {\bibfnamefont {S.}~\bibnamefont {Carniani}}, \bibinfo {author} {\bibfnamefont {S.}~\bibnamefont {Charlot}}, \bibinfo {author} {\bibfnamefont {J.}~\bibnamefont {Chevallard}}, \bibinfo {author} {\bibfnamefont {M.}~\bibnamefont {Curti}}, \bibinfo {author} {\bibfnamefont {E.}~\bibnamefont {Curtis-Lake}}, \bibinfo {author} {\bibfnamefont {F.}~\bibnamefont {D'Eugenio}}, \bibinfo {author} {\bibfnamefont {P.}~\bibnamefont {Jakobsen}}, \bibinfo {author} {\bibfnamefont {T.~J.}\ \bibnamefont {Looser}}, \bibinfo {author} {\bibfnamefont {N.}~\bibnamefont {L{\"u}tzgendorf}}, \bibinfo {author} {\bibfnamefont {R.}~\bibnamefont {Maiolino}},
  \bibinfo {author} {\bibfnamefont {M.~V.}\ \bibnamefont {Maseda}}, \bibinfo {author} {\bibfnamefont {T.}~\bibnamefont {Rawle}}, \bibinfo {author} {\bibfnamefont {H.-W.}\ \bibnamefont {Rix}}, \bibinfo {author} {\bibfnamefont {R.}~\bibnamefont {Smit}}, \bibinfo {author} {\bibfnamefont {H.}~\bibnamefont {{\"U}bler}}, \bibinfo {author} {\bibfnamefont {C.}~\bibnamefont {Willott}}, \bibinfo {author} {\bibfnamefont {J.}~\bibnamefont {Witstok}}, \bibinfo {author} {\bibfnamefont {S.}~\bibnamefont {Baum}}, \bibinfo {author} {\bibfnamefont {R.}~\bibnamefont {Bhatawdekar}}, \bibinfo {author} {\bibfnamefont {K.}~\bibnamefont {Boyett}}, \bibinfo {author} {\bibfnamefont {Z.}~\bibnamefont {Chen}}, \bibinfo {author} {\bibfnamefont {A.}~\bibnamefont {de~Graaff}}, \bibinfo {author} {\bibfnamefont {M.}~\bibnamefont {Florian}}, \bibinfo {author} {\bibfnamefont {J.~M.}\ \bibnamefont {Helton}}, \bibinfo {author} {\bibfnamefont {R.~E.}\ \bibnamefont {Hviding}}, \bibinfo {author} {\bibfnamefont {Z.}~\bibnamefont {Ji}}, \bibinfo
  {author} {\bibfnamefont {N.}~\bibnamefont {Kumari}}, \bibinfo {author} {\bibfnamefont {J.}~\bibnamefont {Lyu}}, \bibinfo {author} {\bibfnamefont {E.}~\bibnamefont {Nelson}}, \bibinfo {author} {\bibfnamefont {L.}~\bibnamefont {Sandles}}, \bibinfo {author} {\bibfnamefont {A.}~\bibnamefont {Saxena}}, \bibinfo {author} {\bibfnamefont {K.~A.}\ \bibnamefont {Suess}}, \bibinfo {author} {\bibfnamefont {F.}~\bibnamefont {Sun}}, \bibinfo {author} {\bibfnamefont {M.}~\bibnamefont {Topping}},\ and\ \bibinfo {author} {\bibfnamefont {I.~E.~B.}\ \bibnamefont {Wallace}},\ }\bibfield  {title} {\bibinfo {title} {Identification and properties of intense star-forming galaxies at redshifts z{\thinspace}>{\thinspace}10},\ }\href {https://doi.org/10.1038/s41550-023-01921-1} {\bibfield  {journal} {\bibinfo  {journal} {Nature Astronomy}\ }\textbf {\bibinfo {volume} {7}},\ \bibinfo {pages} {611} (\bibinfo {year} {2023})}\BibitemShut {NoStop}%
\bibitem [{\citenamefont {Mandel}\ and\ \citenamefont {Wolf}(1995)}]{MandelWolf1995}%
  \BibitemOpen
  \bibfield  {author} {\bibinfo {author} {\bibfnamefont {L.}~\bibnamefont {Mandel}}\ and\ \bibinfo {author} {\bibfnamefont {E.}~\bibnamefont {Wolf}},\ }\href@noop {} {\emph {\bibinfo {title} {Optical Coherence and Quantum Optics}}}\ (\bibinfo  {publisher} {Cambridge University Press},\ \bibinfo {year} {1995})\BibitemShut {NoStop}%
\bibitem [{\citenamefont {Chernoff}(1952)}]{Chernoff1952}%
  \BibitemOpen
  \bibfield  {author} {\bibinfo {author} {\bibfnamefont {H.}~\bibnamefont {Chernoff}},\ }\bibfield  {title} {\bibinfo {title} {{A Measure of Asymptotic Efficiency for Tests of a Hypothesis Based on the sum of Observations}},\ }\href {https://doi.org/10.1214/aoms/1177729330} {\bibfield  {journal} {\bibinfo  {journal} {The Annals of Mathematical Statistics}\ }\textbf {\bibinfo {volume} {23}},\ \bibinfo {pages} {493 } (\bibinfo {year} {1952})}\BibitemShut {NoStop}%
\bibitem [{\citenamefont {Lloyd}\ \emph {et~al.}(2014)\citenamefont {Lloyd}, \citenamefont {Mohseni},\ and\ \citenamefont {Rebentrost}}]{Lloyd2014}%
  \BibitemOpen
  \bibfield  {author} {\bibinfo {author} {\bibfnamefont {S.}~\bibnamefont {Lloyd}}, \bibinfo {author} {\bibfnamefont {M.}~\bibnamefont {Mohseni}},\ and\ \bibinfo {author} {\bibfnamefont {P.}~\bibnamefont {Rebentrost}},\ }\bibfield  {title} {\bibinfo {title} {Quantum principal component analysis},\ }\href {https://doi.org/10.1038/nphys3029} {\bibfield  {journal} {\bibinfo  {journal} {Nature Physics}\ }\textbf {\bibinfo {volume} {10}},\ \bibinfo {pages} {631} (\bibinfo {year} {2014})}\BibitemShut {NoStop}%
\bibitem [{\citenamefont {Khabiboulline}\ \emph {et~al.}(2019)\citenamefont {Khabiboulline}, \citenamefont {Borregaard}, \citenamefont {De~Greve},\ and\ \citenamefont {Lukin}}]{Khabiboulline2019}%
  \BibitemOpen
  \bibfield  {author} {\bibinfo {author} {\bibfnamefont {E.~T.}\ \bibnamefont {Khabiboulline}}, \bibinfo {author} {\bibfnamefont {J.}~\bibnamefont {Borregaard}}, \bibinfo {author} {\bibfnamefont {K.}~\bibnamefont {De~Greve}},\ and\ \bibinfo {author} {\bibfnamefont {M.~D.}\ \bibnamefont {Lukin}},\ }\bibfield  {title} {\bibinfo {title} {Optical interferometry with quantum networks},\ }\href {https://doi.org/10.1103/PhysRevLett.123.070504} {\bibfield  {journal} {\bibinfo  {journal} {Phys. Rev. Lett.}\ }\textbf {\bibinfo {volume} {123}},\ \bibinfo {pages} {070504} (\bibinfo {year} {2019})}\BibitemShut {NoStop}%
\bibitem [{\citenamefont {Mokeev}\ \emph {et~al.}(2025)\citenamefont {Mokeev}, \citenamefont {Saif}, \citenamefont {Lukin},\ and\ \citenamefont {Borregaard}}]{Mokeev2025}%
  \BibitemOpen
  \bibfield  {author} {\bibinfo {author} {\bibfnamefont {A.}~\bibnamefont {Mokeev}}, \bibinfo {author} {\bibfnamefont {B.}~\bibnamefont {Saif}}, \bibinfo {author} {\bibfnamefont {M.~D.}\ \bibnamefont {Lukin}},\ and\ \bibinfo {author} {\bibfnamefont {J.}~\bibnamefont {Borregaard}},\ }\href {https://arxiv.org/abs/2509.09465} {\bibinfo {title} {Enhancing optical imaging via quantum computation}} (\bibinfo {year} {2025}),\ \Eprint {https://arxiv.org/abs/2509.09465} {arXiv:2509.09465 [quant-ph]} \BibitemShut {NoStop}%
\bibitem [{\citenamefont {Berta}\ \emph {et~al.}(2021)\citenamefont {Berta}, \citenamefont {Brand{\~a}o},\ and\ \citenamefont {Hirche}}]{Berta2021}%
  \BibitemOpen
  \bibfield  {author} {\bibinfo {author} {\bibfnamefont {M.}~\bibnamefont {Berta}}, \bibinfo {author} {\bibfnamefont {F.~G. S.~L.}\ \bibnamefont {Brand{\~a}o}},\ and\ \bibinfo {author} {\bibfnamefont {C.}~\bibnamefont {Hirche}},\ }\bibfield  {title} {\bibinfo {title} {On composite quantum hypothesis testing},\ }\href {https://doi.org/10.1007/s00220-021-04133-8} {\bibfield  {journal} {\bibinfo  {journal} {Communications in Mathematical Physics}\ }\textbf {\bibinfo {volume} {385}},\ \bibinfo {pages} {55} (\bibinfo {year} {2021})}\BibitemShut {NoStop}%
\bibitem [{\citenamefont {Lami}(2025)}]{Lami2025}%
  \BibitemOpen
  \bibfield  {author} {\bibinfo {author} {\bibfnamefont {L.}~\bibnamefont {Lami}},\ }\href {https://arxiv.org/abs/2510.06340} {\bibinfo {title} {Generalised quantum sanov theorem revisited}} (\bibinfo {year} {2025}),\ \Eprint {https://arxiv.org/abs/2510.06340} {arXiv:2510.06340 [quant-ph]} \BibitemShut {NoStop}%
\end{thebibliography}%

\end{document}